\documentclass[aps,preprint,tightenlines,superscriptaddress,nofootinbib,showpacs,showkeys]{revtex4}
\usepackage{float}
\usepackage{mathrsfs}
\usepackage{amsfonts}
\usepackage{array}
\usepackage{epsfig}
\usepackage{amsmath}    % need for subequations
\usepackage{amssymb}   % defines \lesssim, etc
\usepackage{graphicx}   % need for figures
\usepackage{verbatim}   % useful for program listings
\usepackage{color}      % use if color is used in text
\usepackage{subfigure}  % use for side-by-side figures
\usepackage{hyperref}
\usepackage{url}
\usepackage{titletoc}
\titlecontents{section}[0pt]{\addvspace{10pt}\filright}
{\contentspush{\thecontentslabel\ }}
{}{\titlerule*[8pt]{.}\contentspage}
\titlecontents{subsection}[2em]{\addvspace{4pt}\filright}
{\contentspush{\thecontentslabel\ }}
{}{\titlerule*[8pt]{.}\contentspage}
\begin{document}

\title{ QCD analysis of CMS $W +$ charm  measurements at LHC with $\sqrt s = 7TeV$
 and implications for strange PDF}

\author{Nijat Yalkun}
%\email{nijat\_phy@aliyun.com}
\affiliation{
School of Physics Science and Technology, Xinjiang University,
 Urumqi, Xinjiang 830046 China }
\affiliation{CAS Key Laboratory of Theoretical Physics, Institute of Theoretical Physics, Chinese Academy of Sciences, Beijing 100190,China}
\affiliation{University of Chinese Academy of Sciences (UCAS), Beijing 100049, China}
\author{Sayipjamal Dulat}
\email{sdulat@hotmail.com}
\affiliation{
School of Physics Science and Technology, Xinjiang University,
 Urumqi, Xinjiang 830046 China }

\begin{abstract}
	We calculate cross-sections and cross-section ratios of a charm quark production in association with a  $W$ gauge boson at next-to-leading order  QCD using MadGraph and CT10NNLO, CT14NNLO, and MSTW2008NNLO PDFs. We compare the results with measurements from the CMS detector at the LHC at a center-of-mass energy of 7TeV.
Moreover, we calculate absolute and normalized differential cross-sections as well as differential cross-section ratios  as a function of the lepton pseudorapidity from the $W$ boson decay.
The correlation  between the CT14NNLO PDFs and predictions for $W+$ charm data  are studied as well. Furthermore, by employing the error PDF updating method  proposed by the CTEQ-TEA group, we update CT14NNLO PDFs, and  analyze the impact of CMS 7TeV $W+$ charm production data to the original CT14NNLO PDFs.
By comparison of the $g(x,Q)$, $s(x,Q)$,  $u(x,Q)$, $d(x,Q)$,  $\bar u(x,Q)$, and $\bar d(x,Q)$ PDFs at $Q=1.3$ GeV and $Q = 100$ GeV for the CT14NNLO and CT14NNLO+Wc, we see that the error band of the $s(x,Q)$  PDF is reduced in the region $x<0.4$, and the error band of $g(x,Q)$ PDF is also slightly reduced in the region $0.01 < x < 0.1$.
\end{abstract}

\maketitle
\thispagestyle{empty}
\newpage
\tableofcontents
\thispagestyle{empty}
\newpage
\setcounter{page}{0}

\section{Introduction} \label{sec:Introduction}

In the standard model (SM), the associated $W +$ charm production in hadron collisions is described at leading order (LO) in perturbative quantum chromodynamics (QCD) by
$g+q \to W^- + c$, $(q=d,s,b)$ and $g+ \bar q\to  W^+ + \bar c$, $(\bar q= \bar d, \bar s, \bar b)$.
Although the d-quark parton distribution function (PDF) is large
in the proton, the processes  $g + d \to W^- + c$ and $g + \bar d \to  W^+ + \bar c$ contribute
only about 10\% \cite{WJ} to the total $W +$ charm production rate, because it is suppressed by the small quark-mixing Cabibbo-Kobayashi-Maskawa (CKM) matrix element\cite{CKM} $|V_{cd}|$ and $|V_{\bar c\bar d}|$.
The major contribution to the total $W +$ charm production rate is due to strange quark-gluon  fusion
$g + s \to W^- + c$, and $g + \bar s \to  W^+ + \bar c$.
The contribution from  $g+ b \to W^- + c$ and $g + \bar b \to  W^+ + \bar c$ is
also heavily suppressed by the quark mixing matrix elements ($|V_{cb}|$, $|V_{\bar c\bar b}|$) and the b-quark PDF.
The $W +$ charm production cross-section is therefore particularly
sensitive to the  proton  $g(x, Q)$ and $s(x, Q)$  PDFs \cite{1U} and to the magnitude of the CKM matrix element $V_{cs}$, where $x$ is the momentum fraction of the proton carried by the s-quark, and $Q$ is the hard scale.
The study cited in Ref.\cite{1K} calculated the $W +$ charm production at LO and next-to-leading order (NLO) in QCD,
and found that the factorization and renormalization scale uncertainty in the NLO calculation is about 20\%.
Ref.\cite{Lai:2007dq}  explored the strangeness degrees of freedom in the parton structure of the nucleon within the global analysis framework, and showed that the precise determination of the $s(x,Q)$ PDF affects the $W +$ charm cross-section.
The $s(x,Q)$ PDF has been determined by neutrino-nucleon deep inelastic scattering experiments at momentum transfer squared $Q^2=10$ GeV, and momentum fraction $ x \sim 0.1$ \cite{NuTeV01, NuTeV07}.
The Tevatron CDF\cite{CDF} and D0\cite{D0} experiments have measured the cross-section for charm quark produced in association with $W$ bosons, using muon tagging of the charm-quark jet. The ATLAS collaboration \cite{atlas7}  measured the total cross-section,
differential cross-section as a function of the pseudorapidity of the lepton from the $W$ boson decay, and the cross-section ratio of the
production of a $W$ boson in association with a single charm quark  at $\sqrt s = 7$ TeV.
The CMS experiment measured \cite{CMS7}  total cross-sections $( \sigma(W^- + c), \sigma(W^+ + \bar{c}))$,  absolute and normalized differential cross-sections as a function of the absolute value of the pseudorapidity of the lepton from the $W$ boson decay, and the  cross-section ratio
$ R^c = \sigma(W^+ + \bar c)/\sigma(W^- + c)$ at a center of mass energy 7TeV  for the fiducial region defined, namely
\begin{eqnarray} \label{eq1}
&& p_T^{j}> 25 ~\text{GeV}, \;\; |\eta^j|<2.5, \;\; |\eta^l|<2.1,\nonumber \\
&& p_T^{l}> 25 ~\text{GeV}, \;\; \text{for} \;\;  W \to \mu \nu_{\mu},\\
&& p_T^{l}> 35 ~\text{GeV}, \;\; \text{for} \;\;  W \to \mu \nu_{\mu}  \;\; \text{and}\;\;
W \to e \nu_{e}.\nonumber
\end{eqnarray}
 There are two different transverse momentum cuts for the charged lepton in the final state. When $p_T^{l}> 25 ~\text{GeV}$, we only consider the muon decay channel($W\to \mu \nu_\mu$) for $W$ boson; both muon($W\to \mu \nu_\mu$) and electron($W\to e \nu_e$) decay channels for $W$ boson are considered, when $p_T^{l}> 35 ~\text{GeV}$.
This study is organized as follows: in Section \ref{sec3}, we present our results for various latest PDF sets and compare these with the CMS measurements of the total cross-section, absolute and normalized differential cross-sections and ratios, as well as the correlation  between the CT14NNLO PDFs and predictions for $W+$ charm data.
In Section \ref{sec4}, we discuss the impact of the CMS $W+$ charm production 7TeV data to the CT14NNLO PDFs.
In Section \ref{sec5}, we draw our conclusions.

\section{Results}\label{sec3}

\begin{figure}[H]
	\begin{center}
    \includegraphics[width=0.95\textwidth]{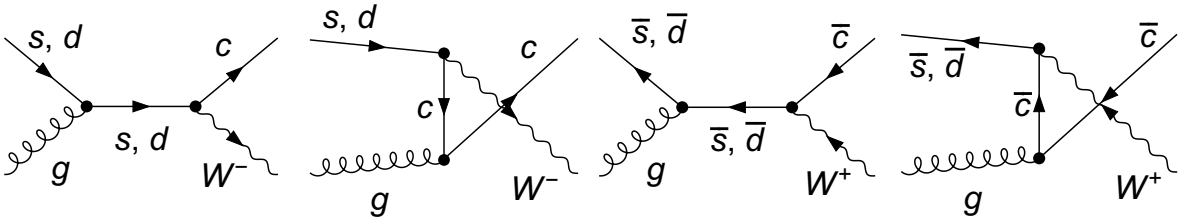}
\caption{ Possible tree-level diagrams at partonic level for $W +$ charm production}
	\label{fig:feynVertices}
	\end{center}
\end{figure}

In this section, we present a detailed numerical study of the $pp \to W + c + X$ process at the LHC at a center-of-mass energy of 7TeV at NLO order QCD using the Monte-Carlo numerical calculation program MadGraph\cite{6} with CT10NNLO\cite{8}, CT14NNLO\cite{9}, and MSTW2008NNLO\cite{10} PDFs.
PDF uncertainties on the theoretical predictions are given at 68\% confidence level (C.L.).
We calculate the total cross-section, differential (absolute and normalized) cross-sections, and the cross-section ratio
$ R^c = \sigma(W^+ + \bar c)/\sigma(W^- + c)$  with the  $W \to l \nu$ decay (where $l= \mu$ or $e$).
In our study, we use the same kinematical cuts as the CMS detector at the LHC at a center-of-mass energy of 7TeV \cite{CMS7}, that are given in  section \ref{sec:Introduction};
both the factorization and renormalization scales
are set to the value of the $W$ boson mass $\mu_R=\mu_F=m_{\text{w}}$;
charm quark mass is considered and is set to 1.550 GeV; strong interaction coupling $\alpha_s$ is set to 0.118, and for  electro-weak parameters, the $W$ boson mass is set to 80.385 GeV; Fermi coupling  is set to as $1.166\times10^{-5}$ $\text{GeV}^{-2}$; related CKM matrix elements are set to as $V^{cd}=0.225$ and $V^{cs}=0.974$; the mass of charged light leptons is considered and set to as $m_e=0.511$ MeV and $m_\mu=105.658$ MeV.
At LO, the Feynman diagrams for the hard scattering processes of the $W+$ charm production $pp\to W +c+X$  are shown in Fig.\ref{fig:feynVertices}.
The main contribution  for the cross-sections of $W+$ charm production comes from strange quark and gluon scattering, the down-quark contribution is strongly Cabibbo suppressed, and the contribution from the bottom quark and gluon scattering is negligible.

\subsection{Total cross-section}\label{tXsec}

\begin{figure}[H]
\begin{center}
\includegraphics[width=0.45\textwidth]{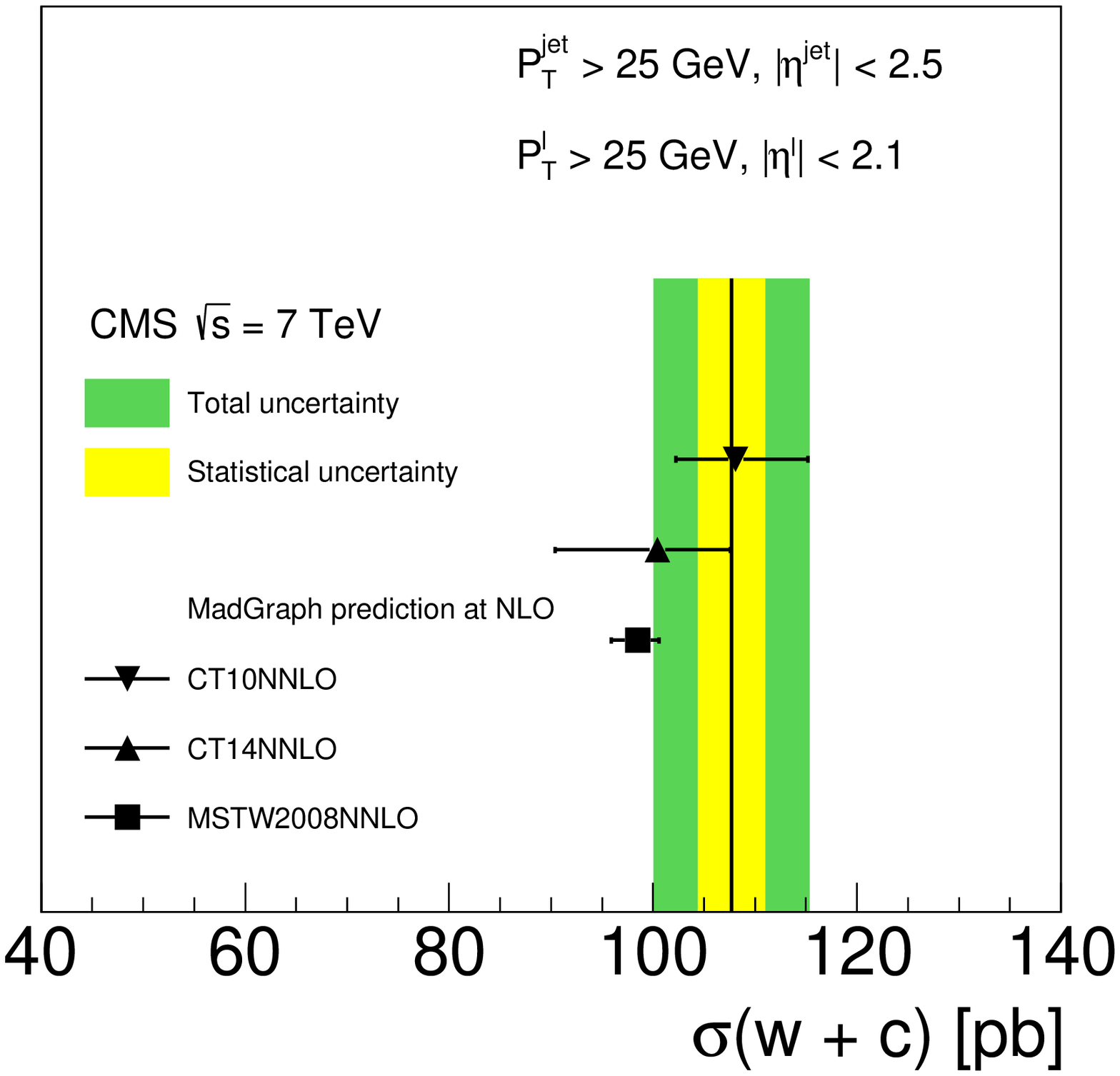}
\includegraphics[width=0.45\textwidth]{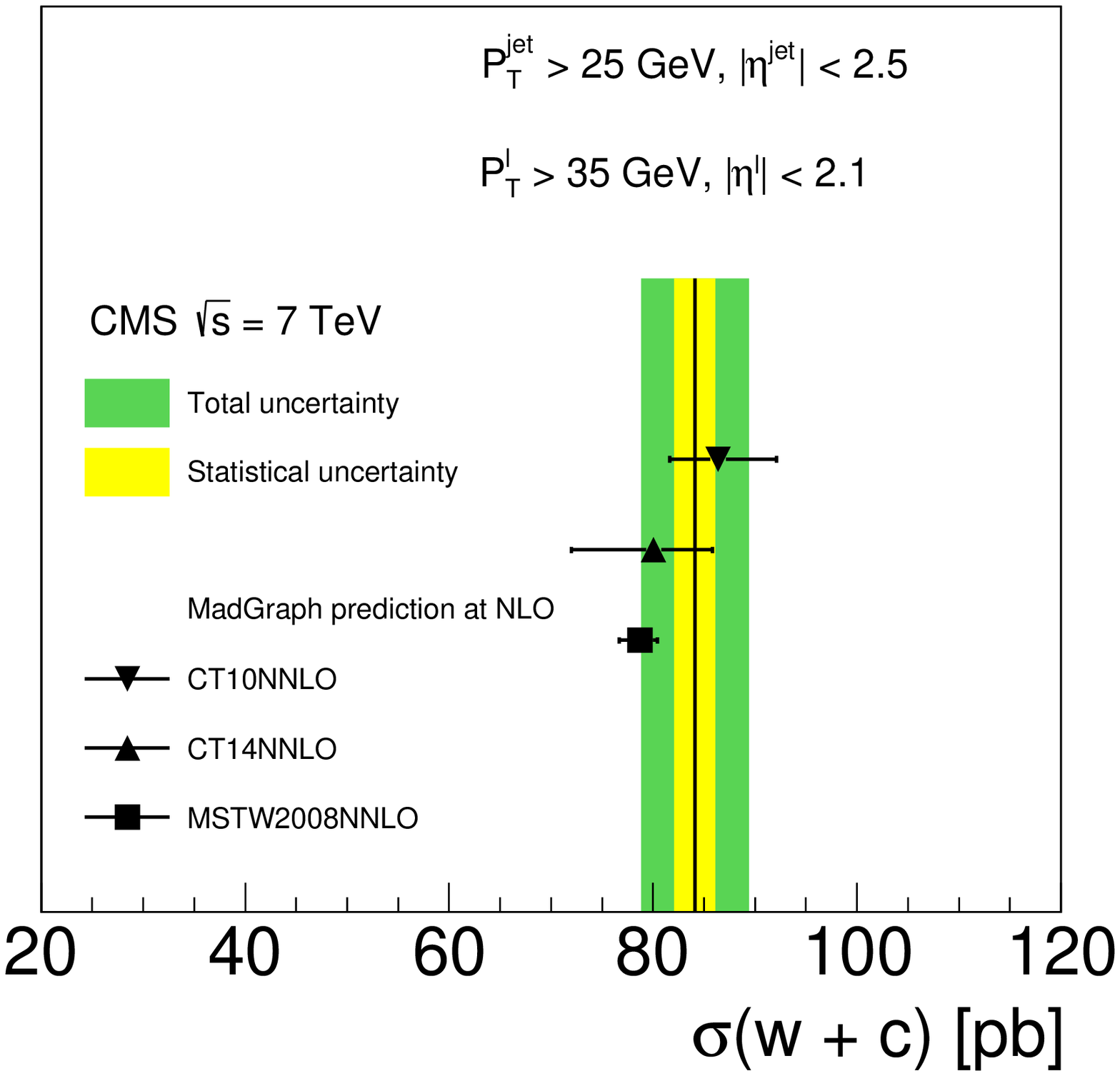}
\caption{Comparison of theoretical predictions for total cross-section $\sigma(W + c)$ computed with MadGraph using the CT10NNLO, CT14NNLO and MSTW2008NNLO PDFs with CMS  measurements. The left figure contains the prediction  for the lepton from $W$ boson decay with $p^l_T > 25 $GeV,  while the right figure is for the case $p^l_T > 35$GeV.
The solid vertical line shows the central value of the measurement, the inner error
band corresponds to the statistical uncertainty, and the outer error band to the sum in quadrature of the statistical and systematic uncertainties.}\label{Totxsecfig}
\end{center}
\end{figure}

The total cross-sections $\sigma(W^+ +\bar c)$ and $\sigma(W^- + c)$
of the production of a $W$ boson in association with a charm quark in pp collisions at
$\sqrt s = 7$ TeV at NLO QCD are summarized in Table \ref{Totxsectab}.
PDF uncertainties  are at 68\% confidence level (C.L.), that are obtained from the error sets of the CT10NNLO, CT14NNLO, and MSTW2008NNLO.
The experimental measurements  from CMS collaboration at the LHC at a center-of-mass energy of 7TeV \cite{CMS7} are also included in this table.
The comparison between theory predictions based on various PDF sets and experimental measurements are  illustrated in Fig \ref{Totxsecfig}.
The predictions obtained with the CT10NNLO PDFs are in agreement with CMS
measurements.
The predictions obtained with CT14NNLO agree with CMS measurements within the uncertainty range.
The prediction obtained with MSTW2008NNLO is less favored.
Those differences in the size of the PDF uncertainties depend on a different methodology
and the  parametrization of the $s(x,Q)$ PDF used by different PDF sets.

\begin{table}[H]
	\centering
	\caption{Total cross-section of $\sigma(pp\to W + c )\times B(W\to\ l \nu)$ }
	\label{Totxsectab}
	\begin{tabular}{ c |c | c }
		\hline\hline
		&\multicolumn{2}{c}{$\sigma( pp \to W + c ) \times B( W \to l \nu )$[pb]}\\
		PDF sets&{$p^l_T>25$GeV} &{$p^l_T>35$GeV}\\
		\hline
		CT10NNLO & 108.1 $^{+6.6\%}_{-5.5\%} $& 86.4 $^{+6.7\%}_{-5.5\%} $\\
		CT14NNLO & 100.4 $^{+7.1\%}_{-10.0\%} $ & 80.1 $^{+7.2\%}_{-10.1\%} $\\
		MSTW2008NNLO & 98.5  $^{+2.1\%}_{-2.6\%} $ & 78.7 $^{+2.1\%}_{-2.6\%} $\\
		\hline\hline
		CMS  & {107.7 $\pm$ 3.1\% (stat.) $\pm$ 6.4\% (syst.)} &{ 84.1 $\pm$ 2.4\% (stat.) $\pm$ 5.8\% (syst.)} \\
		\hline
	\end{tabular}	
\end{table}

Our MadGraph calculations of the  $d$, $\bar d$, $s$,  and $\bar s$ quarks contributions (in pb) to the LO $W + $ charm total cross-sections with NNLO PDFs for the leptonic decay channel $W \to  e\nu$ are shown in Table \ref{LOcon}.
The strange quark contributes the most to this $W+$ charm production. With regard to the parametrization of the strange-quark content of the proton, CTEQ-TEA and MSTW2008 PDF groups make different assumptions in their global fits.
In CT10NNLO and CT14NNLO, the strange  is parameterized symmetrically  $s=\bar s$, and in MSTW2008, it is parameterized asymmetrically, $s - \bar s \neq 0$.
Hence corresponding theoretical predictions  differ accordingly.
For MSTW2008, the production of $W^+ + c$   is  slightly larger than the $W^+ + \bar c$,
as expected because of the  $s-\bar s$ asymmetry.
Because of the dominance of the $d$ quark over the $\bar d$ -quark in the proton,
the production of $W^- + c$  is  larger than $W^+ + \bar c$.
The numbers in bracket correspond to $p_T^{l}>35 $ GeV.

\begin{table}[H]
\centering
\caption{LO contributions of  $d$, $\bar d$, $s$  and $\bar s$ quarks to $\sigma(pp\to W + c )\times B(W\to\ l \nu)$ within the kinematic region $p_T^{jet}>25$ GeV, $|\eta^{jet}|<2.5$, lepton pseudorapidity range $|\eta_l|<2.1$  and $p_T^{l}>25(35)$ GeV.}\label{LOcon}
\begin{tabular}{c | c | c | c }
		\hline\hline
		Subprocess&CT10NNLO&CT14NNLO&MSTW2008NNLO\\
		\hline
		$\bar s + g \to W^+ + \bar c $ &35.82(28.72)&32.85(26.28)&31.59(25.49)\\		
		$\bar d + g \to W^+ + \bar c $ &2.33(1.89)&2.37(1.92)&2.43(1.96)\\
		$s + g \to W^- + c $ &35.85(28.78)&32.89(26.32)&32.49(26.15)\\		
		$d + g \to W^- + c $ &4.50(3.73)&4.58(3.78)&4.66(3.86)\\		
		\hline		
\end{tabular}
\end{table}

\subsection{Absolute and normalized  differential cross-section}\label{diffXsec}

\begin{figure}[H]\vspace{-0.1in}
	\begin{center}
		\includegraphics[width=0.45\textwidth]{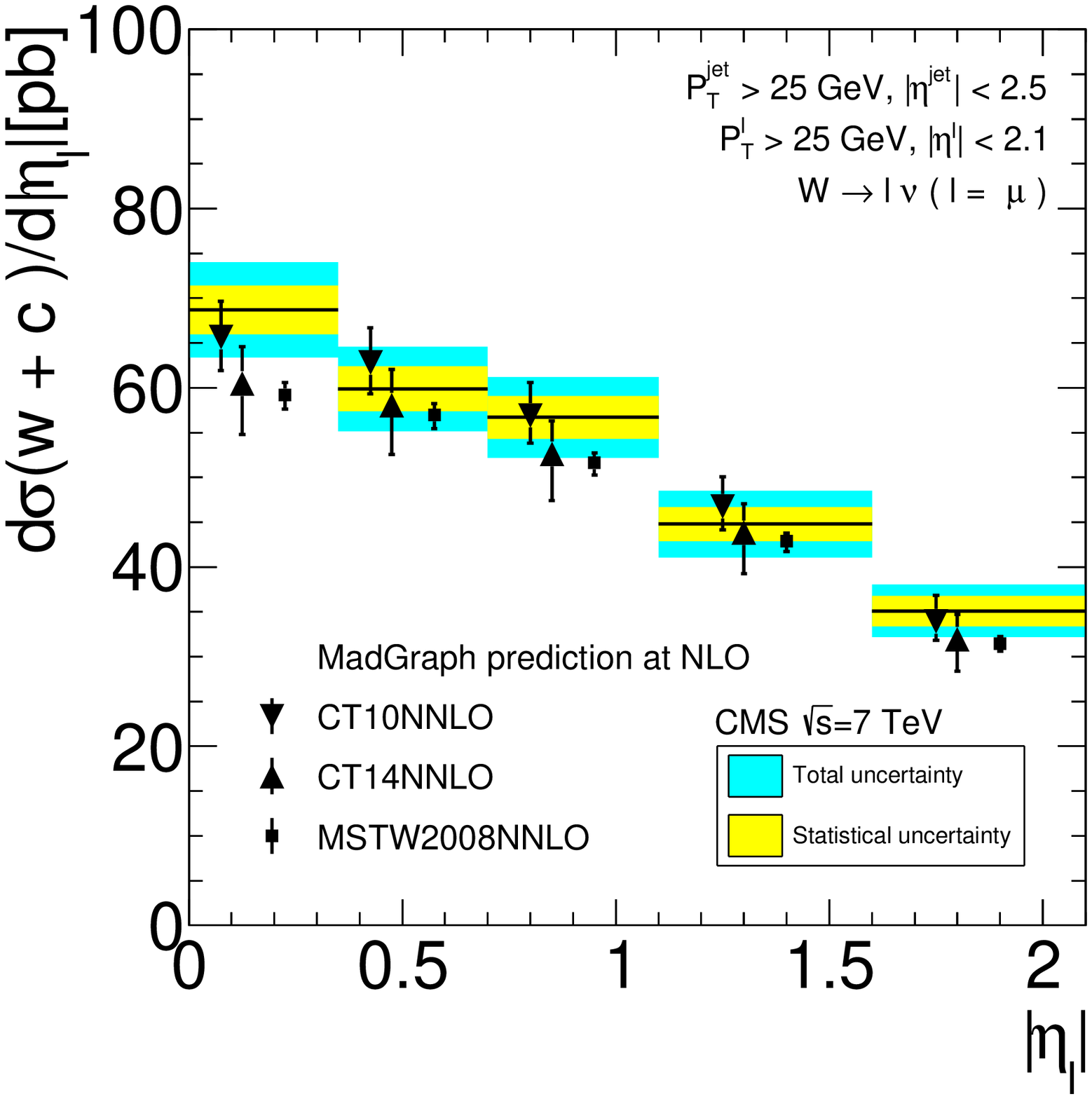}
		\includegraphics[width=0.45\textwidth]{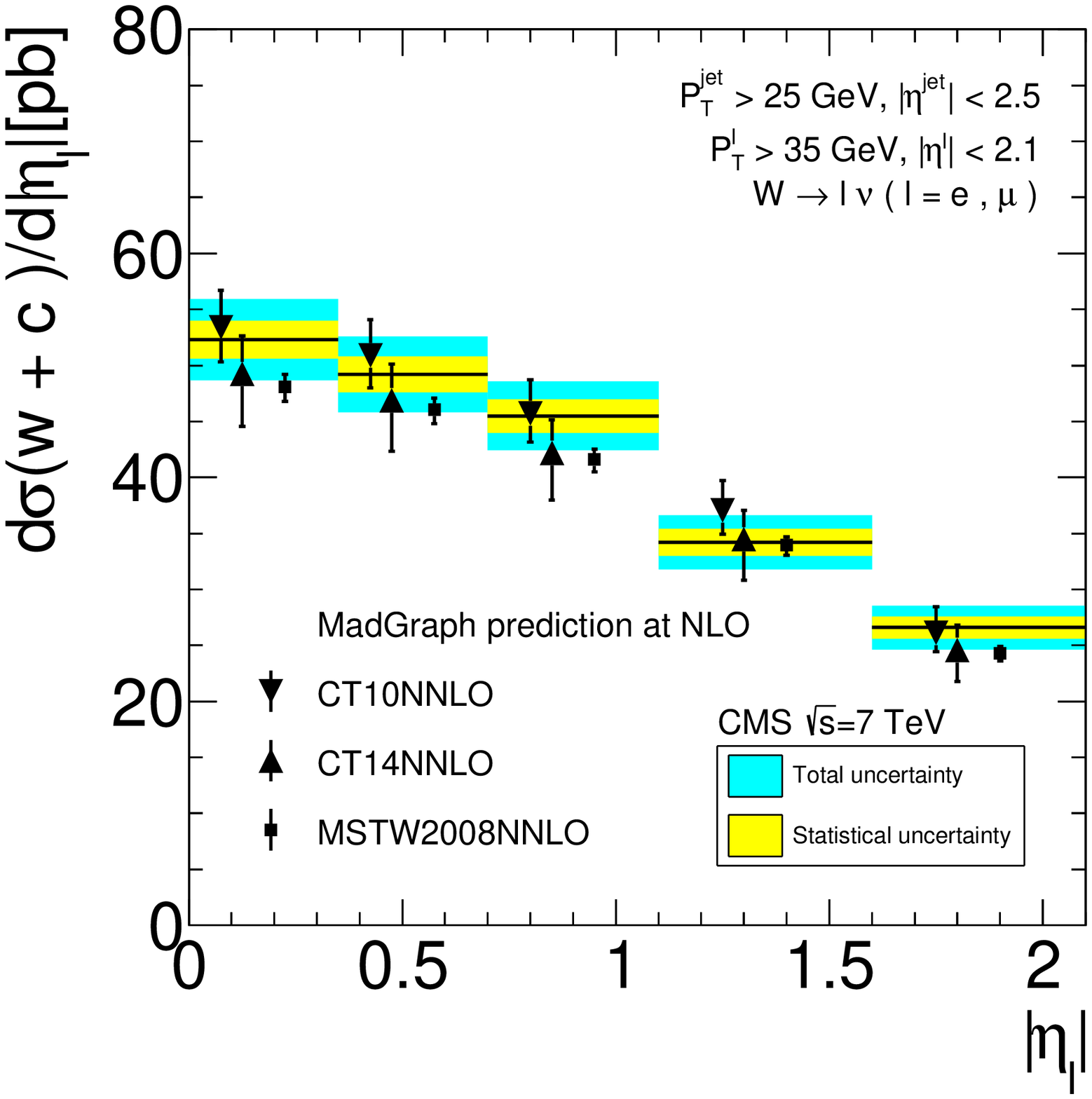}
		\caption{
			Comparison of  theoretical predictions for differential cross-sections, $d\sigma(W+c)/d|\eta|$, as a function of absolute value of pseudorapidity of lepton from W-boson decay, with CMS  measurements.
Theoretical predictions at NLO are calculated using MadGraph with CT10NNLO, CT14NNLO and MSTW2008NNLO PDFs. The left figure shows the predictions  for the lepton from $W$ boson decay with $p^l_T > 25 $GeV, and the right with $p^l_T > 35$GeV. Error bars on the theoretical predictions show 68\% C.L..
		}
		\label{diffxsecfig}
	\end{center}
\end{figure}

\begin{figure}[H]\vspace{-0.1in}
	\begin{center}
		\includegraphics[width=0.45\textwidth]{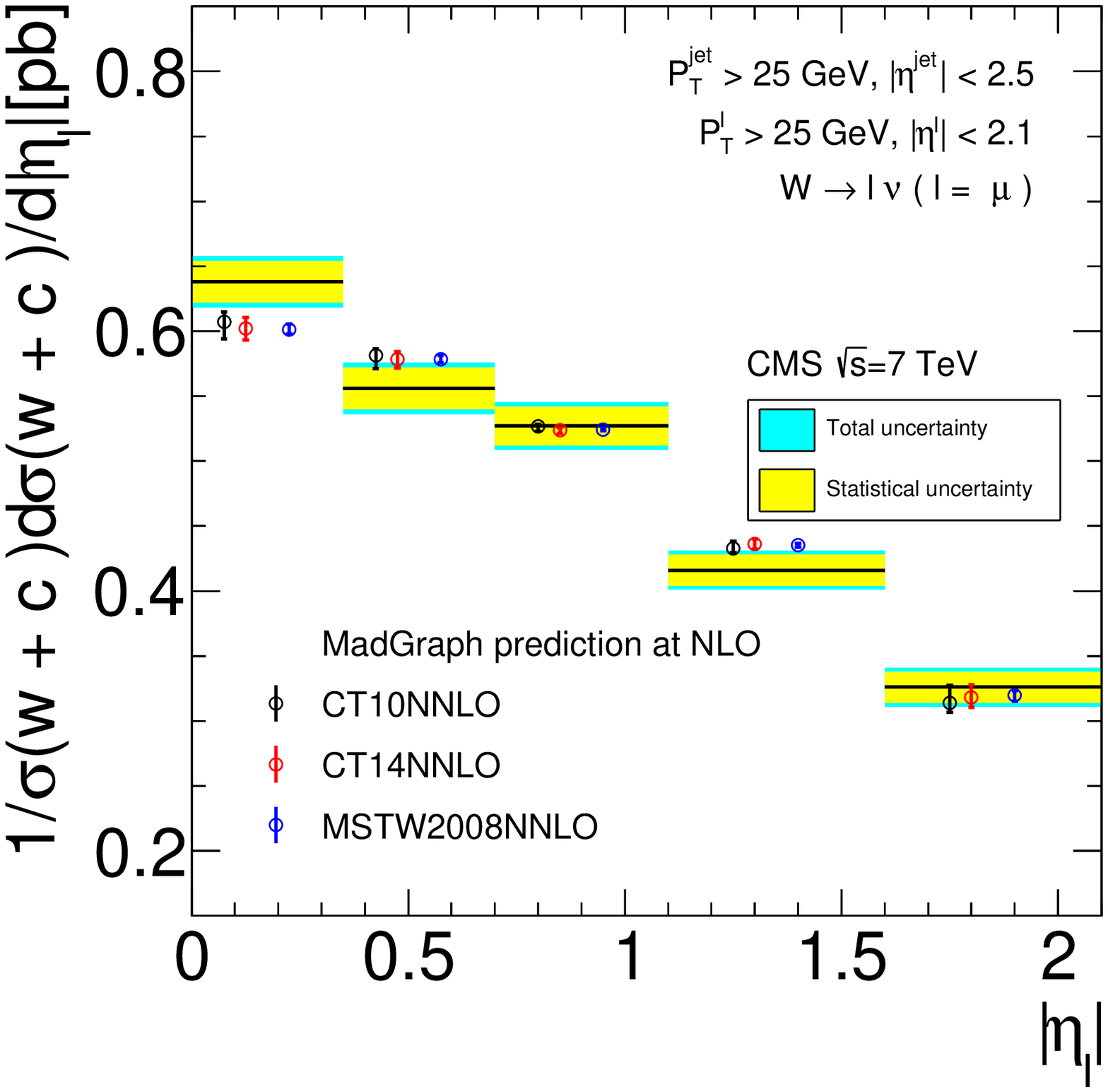}
		\includegraphics[width=0.45\textwidth]{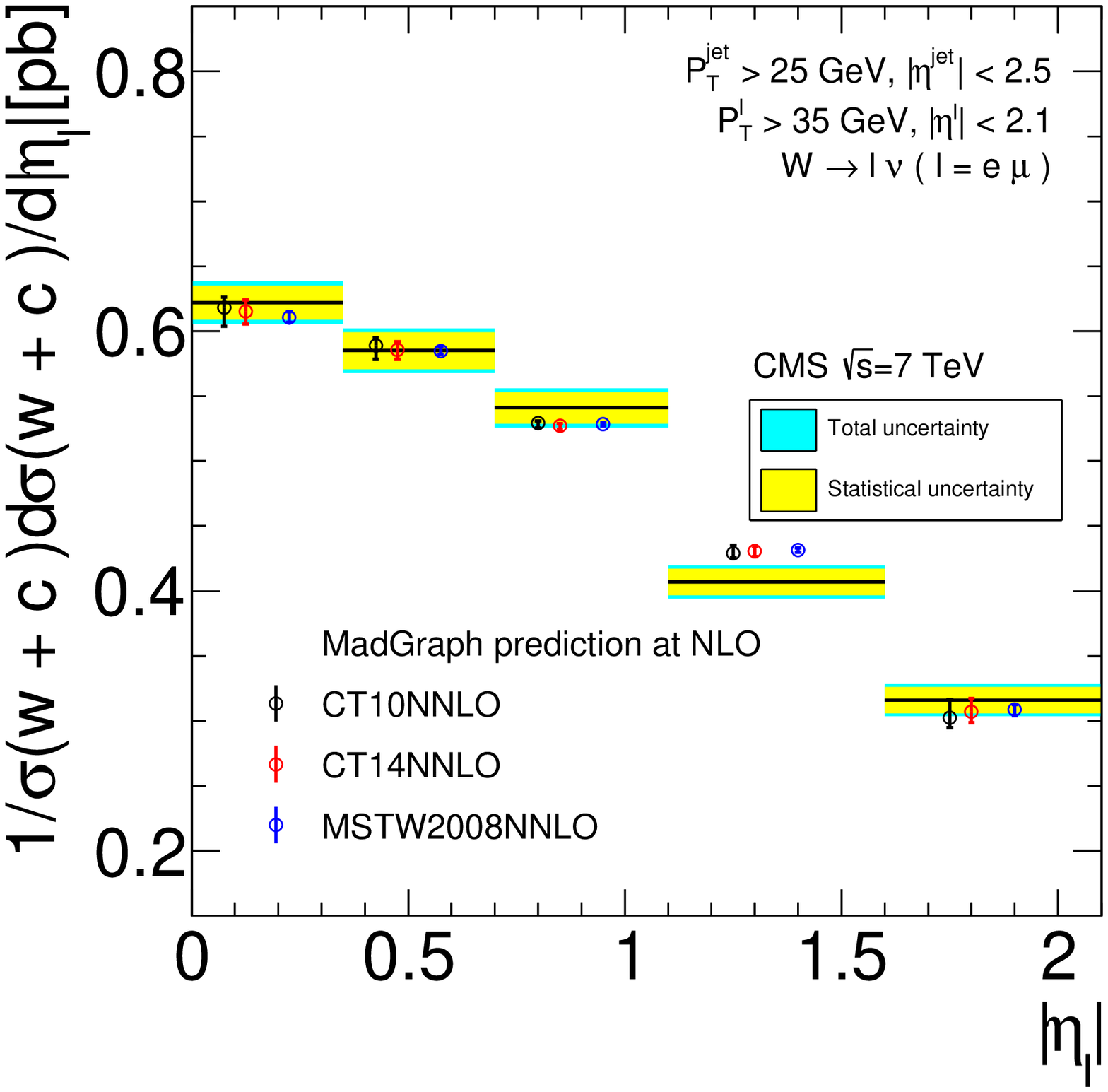}
		\caption{Comparison of theoretical predictions for normalized differential cross-sections, $d\sigma(W+c)/\sigma d|\eta|$, as a function of absolute value of pseudorapidity of lepton from W-boson decay, with CMS  measurements.
Theoretical predictions at NLO are calculated using MadGraph with CT10NNLO, CT14NNLO and MSTW2008NNLO PDFs.
The left figure shows the predictions  for the lepton from the $W$ boson decay with $p^l_T > 25 $GeV, and the right with $p^l_T > 35$GeV.}
		\label{normdiffxsecfig}
	\end{center}
\end{figure}

The absolute and normalized differential cross-sections are obtained by MadGraph using the same setup as the CMS collaboration at the LHC and a center-of-mass energy of 7TeV.
In Fig.\ref{diffxsecfig} and Fig.\ref{normdiffxsecfig}, we compare the absolute and normalized differential  cross-sections in bins of lepton pseudo-rapidity with CMS measurements. The absolute and normalized differential cross-sections with PDF uncertainty at 68\% C.L. are summarized in  Table \ref{diffxsectab} and \ref{normdiffxsectab}, and the CMS 7TeV measurement with statistical and systematic uncertainty is given in the last column.
There is good agreement between theoretical predictions and measured distributions. The comparisons among predictions from various PDFs may lead to different conclusions. For instance, the predictions based on CT14NNLO and MSTW2008NNLO PDFs are smaller than the predictions based on CT10NNLO PDFs, and PDF uncertainties of CT14NNLO PDFs are much larger then the PDF uncertainties of CT10NNLO and MSTW2008NNLO.

\begin{table}[H]
	\centering
	\caption{Theory predictions of differential cross-sections $ d \sigma( pp \to W + c ) \times B( W \to l \nu )/ d |\eta^l| $  with PDF uncertainty at 68 \% C.L., last three column for CMS 7TeV measurement with statistical and systematic uncertainty.}\label{diffxsectab}
	\begin{tabular}{ c | c c c | c }
		\hline\hline
		\multicolumn{5}{c}{$p_T^l>25$GeV}\\
		$ |\eta^l|$ & CT10NNLO & CT14NNLO & MSTW2008NNLO& CMS measurement\\
		\hline
		$ [0,0.35]  $ &$65.7^{+6.1\%}_{-5.7\%}$&$60.5^{+6.8\%}_{-9.4\%}$&$59.2^{+2.3\%}_{-2.7\%}$&$68.7\pm3.9\%\pm6.7\%$\\
		$ [0.35,0.7]$ &$62.8^{+6.2\%}_{-5.6\%}$&$58.1^{+6.9\%}_{-9.5\%}$&$57.0^{+2.3\%}_{-2.7\%}$&$59.9\pm4.2\%\pm6.7\%$\\
		$ [0.7,1.1] $ &$56.9^{+6.4\%}_{-5.5\%}$&$52.6^{+7.0\%}_{-9.8\%}$&$51.6^{+2.2\%}_{-2.6\%}$&$56.7\pm4.2\%\pm6.7\%$\\
		$ [1.1,1.6] $ &$46.8^{+7.0\%}_{-5.7\%}$&$43.8^{+7.4\%}_{-10.4\%}$&$42.9^{+2.2\%}_{-2.7\%}$&$44.8\pm4.2\%\pm7.1\%$\\
		$ [1.6,2.1] $ &$33.9^{+8.5\%}_{-6.2\%}$&$32.0^{+8.7\%}_{-11.3\%}$&$31.5^{+2.4\%}_{-2.9\%}$&$35.1\pm4.8\%\pm6.8\%$\\
		\hline\hline
		\multicolumn{5}{c}{$p_T^l>35$GeV}\\
		$ |\eta^l|$ & CT10NNLO & CT14NNLO & MSTW2008NNLO& CMS measurement\\
		\hline
		$ [0,0.35]  $ &$53.4^{+6.2\%}_{-5.8\%}$&$49.2^{+6.9\%}_{-9.5\%}$&$48.1^{+2.3\%}_{-2.7\%}$&$52.3\pm3.3\%\pm6.1\%$\\
		$ [0.35,0.7]$ &$50.9^{+6.3\%}_{-5.7\%}$&$46.9^{+6.9\%}_{-9.7\%}$&$46.1^{+2.3\%}_{-2.7\%}$&$49.2\pm3.3\%\pm6.1\%$\\
		$ [0.7,1.1] $ &$45.7^{+6.6\%}_{-5.6\%}$&$42.2^{+7.0\%}_{-10.0\%}$&$41.6^{+2.2\%}_{-2.6\%}$&$45.5\pm3.3\%\pm5.9\%$\\
		$ [1.1,1.6] $ &$37.1^{+7.2\%}_{-5.7\%}$&$34.5^{+7.5\%}_{-10.6\%}$&$34.0^{+2.2\%}_{-2.7\%}$&$34.2\pm3.5\%\pm6.1\%$\\
		$ [1.6,2.1] $ &$26.1^{+8.9\%}_{-6.3\%}$&$24.6^{+9.0\%}_{-11.4\%}$&$24.3^{+2.4\%}_{-3.0\%}$&$26.6\pm3.8\%\pm6.4\%$\\
		\hline
		\hline
	\end{tabular}	
\end{table}

\begin{table}[H]	
	\centering
	\caption{Theory predictions of normalized differential cross-sections $ (1/\sigma(W+c))  d\sigma(W+c)/d|\eta|$ with PDF uncertainty at 68 \% C.L., three column depicts CMS 7TeV measurement with statistical and systematic uncertainty.}\label{normdiffxsectab}
	\begin{tabular}{ c | c c c | c }
		\hline\hline
		\multicolumn{5}{c}{$p_T^l>25$GeV}\\
		$ |\eta^l|$ & CT10NNLO & CT14NNLO & MSTW2008NNLO& CMS measurement\\
		\hline
		$ [0,0.35]  $ &$0.607^{+1.2\%}_{-2.2\%}$&$0.602^{+1.4\%}_{-1.5\%}$&$0.601^{+0.7\%}_{-0.6\%}$&$0.638\pm2.5\%\pm1.9\%$\\
		$ [0.35,0.7]$ &$0.581^{+0.9\%}_{-1.7\%}$&$0.578^{+1.1\%}_{-1.2\%}$&$0.578^{+0.5\%}_{-0.4\%}$&$0.556\pm2.9\%\pm2.2\%$\\
		$ [0.7,1.1] $ &$0.527^{+0.3\%}_{-0.7\%}$&$0.524^{+0.4\%}_{-0.5\%}$&$0.524^{+0.2\%}_{-0.2\%}$&$0.527\pm2.8\%\pm2.1\%$\\
		$ [1.1,1.6] $ &$0.433^{+1.3\%}_{-0.8\%}$&$0.436^{+0.8\%}_{-0.9\%}$&$0.435^{+0.3\%}_{-0.4\%}$&$0.416\pm2.9\%\pm2.2\%$\\
		$ [1.6,2.1] $ &$0.314^{+4.2\%}_{-2.3\%}$&$0.318^{+3.0\%}_{-2.5\%}$&$0.320^{+1.1\%}_{-1.3\%}$&$0.326\pm3.7\%\pm2.8\%$\\
		\hline\hline
		\multicolumn{5}{c}{$p_T^l>35$GeV}\\
		$ |\eta^l|$ & CT10NNLO & CT14NNLO & MSTW2008NNLO& CMS measurement\\
		\hline
		$ [0,0.35]  $ &$0.618^{+1.3\%}_{-2.3\%}$&$0.615^{+1.5\%}_{-1.6\%}$&$0.611^{+0.7\%}_{-0.6\%}$&$0.622\pm2.1\%\pm1.6\%$\\
		$ [0.35,0.7]$ &$0.589^{+1.0\%}_{-1.8\%}$&$0.586^{+1.1\%}_{-1.3\%}$&$0.585^{+0.6\%}_{-0.5\%}$&$0.585\pm2.4\%\pm1.7\%$\\
		$ [0.7,1.1] $ &$0.529^{+0.4\%}_{-0.7\%}$&$0.527^{+0.4\%}_{-0.5\%}$&$0.528^{+0.3\%}_{-0.2\%}$&$0.541\pm2.2\%\pm1.7\%$\\
		$ [1.1,1.6] $ &$0.429^{+1.4\%}_{-0.8\%}$&$0.431^{+0.9\%}_{-1.0\%}$&$0.432^{+0.4\%}_{-0.4\%}$&$0.407\pm2.5\%\pm2.0\%$\\
		$ [1.6,2.1] $ &$0.302^{+4.7\%}_{-2.5\%}$&$0.307^{+3.4\%}_{-2.7\%}$&$0.309^{+1.3\%}_{-1.5\%}$&$0.316\pm3.2\%\pm2.2\%$\\
		\hline
		\hline
	\end{tabular}
\end{table}

\subsection{Charged cross-section ratio}\label{ratio}

\begin{figure}[H]
\begin{center}
		\includegraphics[width=0.42\textwidth]{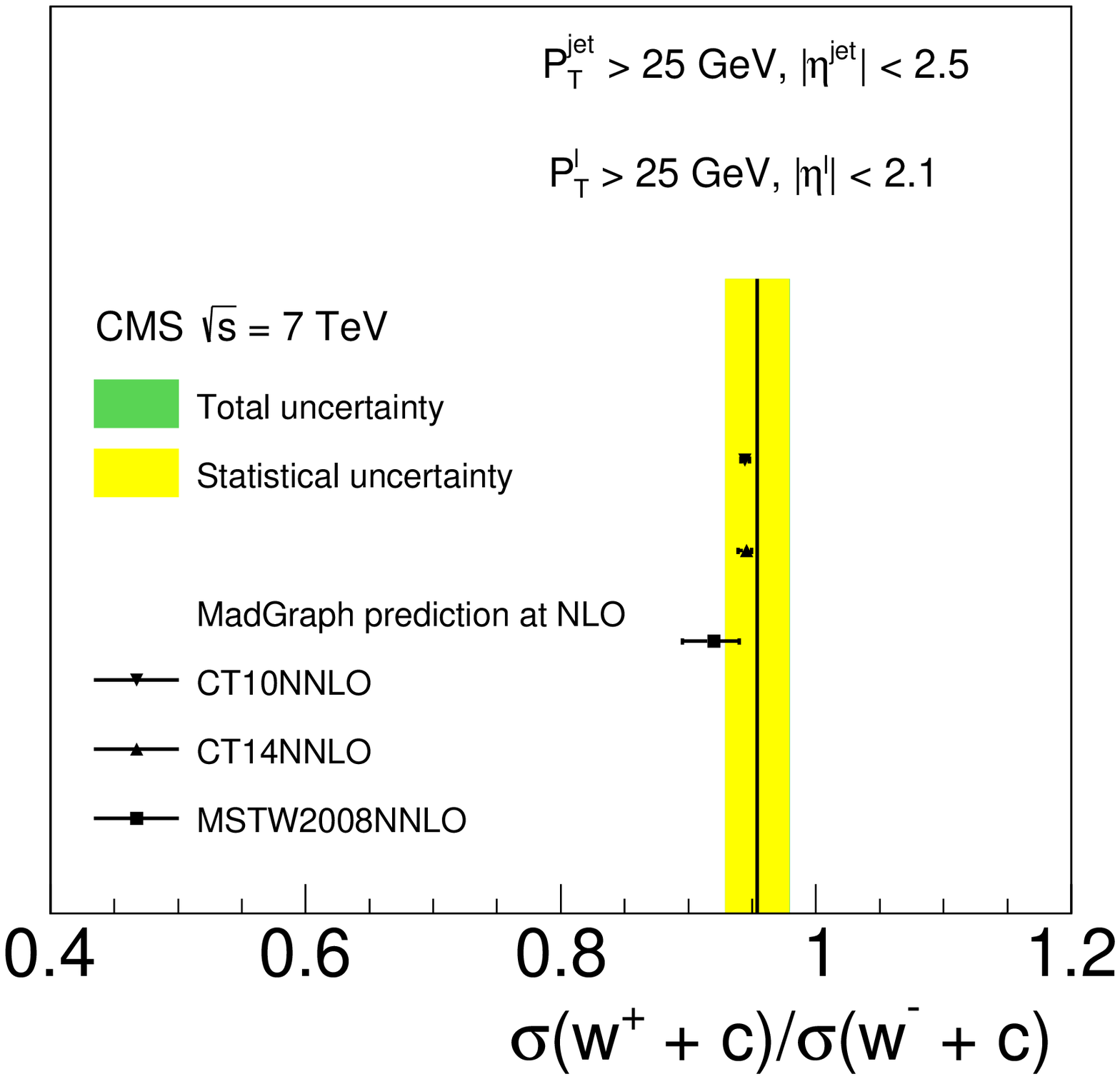}\hspace{10pt}
		\includegraphics[width=0.42\textwidth]{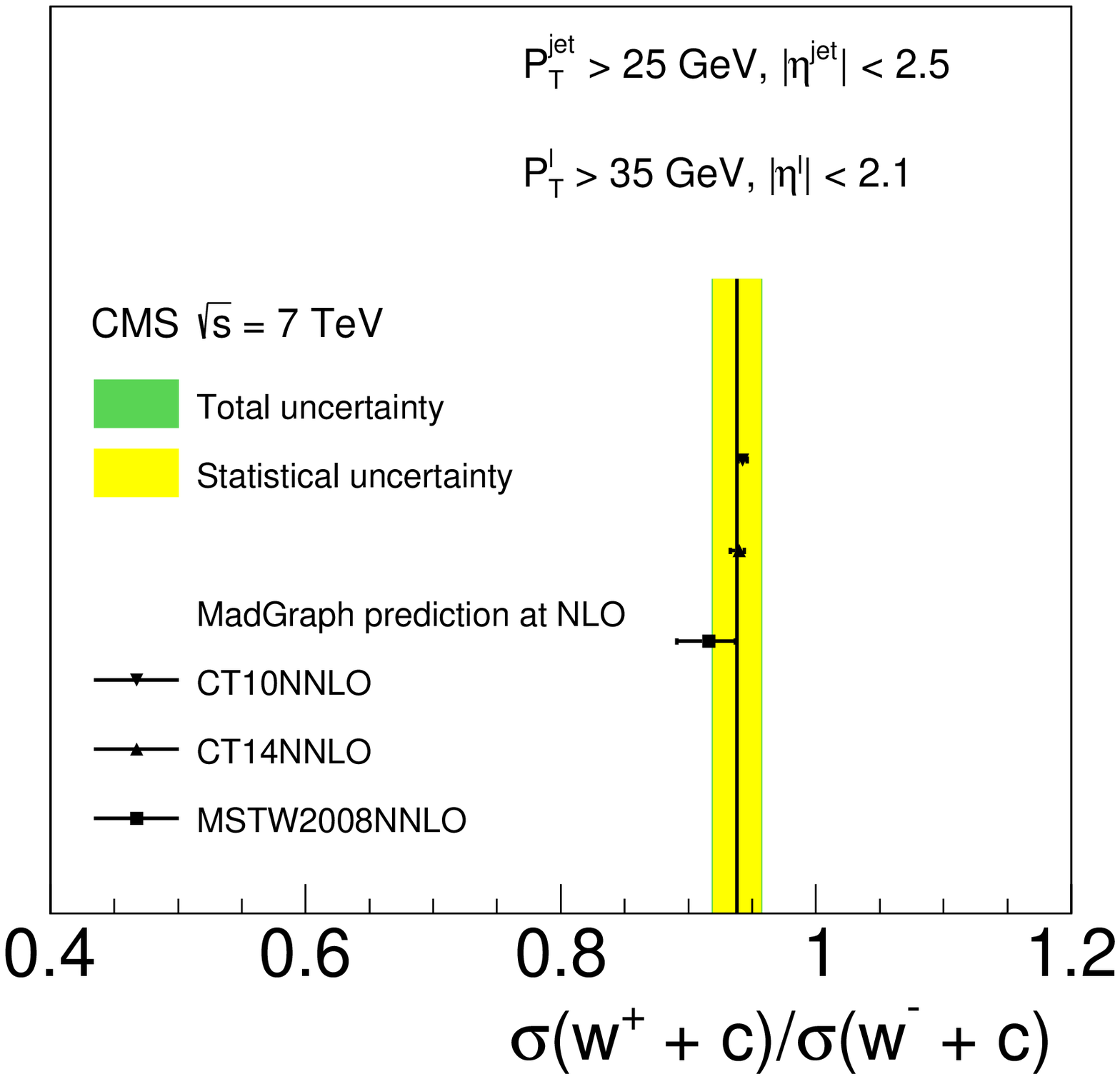}
\caption{Comparison of theoretical prediction of total cross-section ratio $\sigma(W^+ + \bar c)/\sigma(W^- + c)$ for three different PDF sets with CMS  measurements. }\label{totalratiofig}
\end{center}
\end{figure}

\begin{figure}[H]
\begin{center}
	\includegraphics[width=0.45\textwidth]{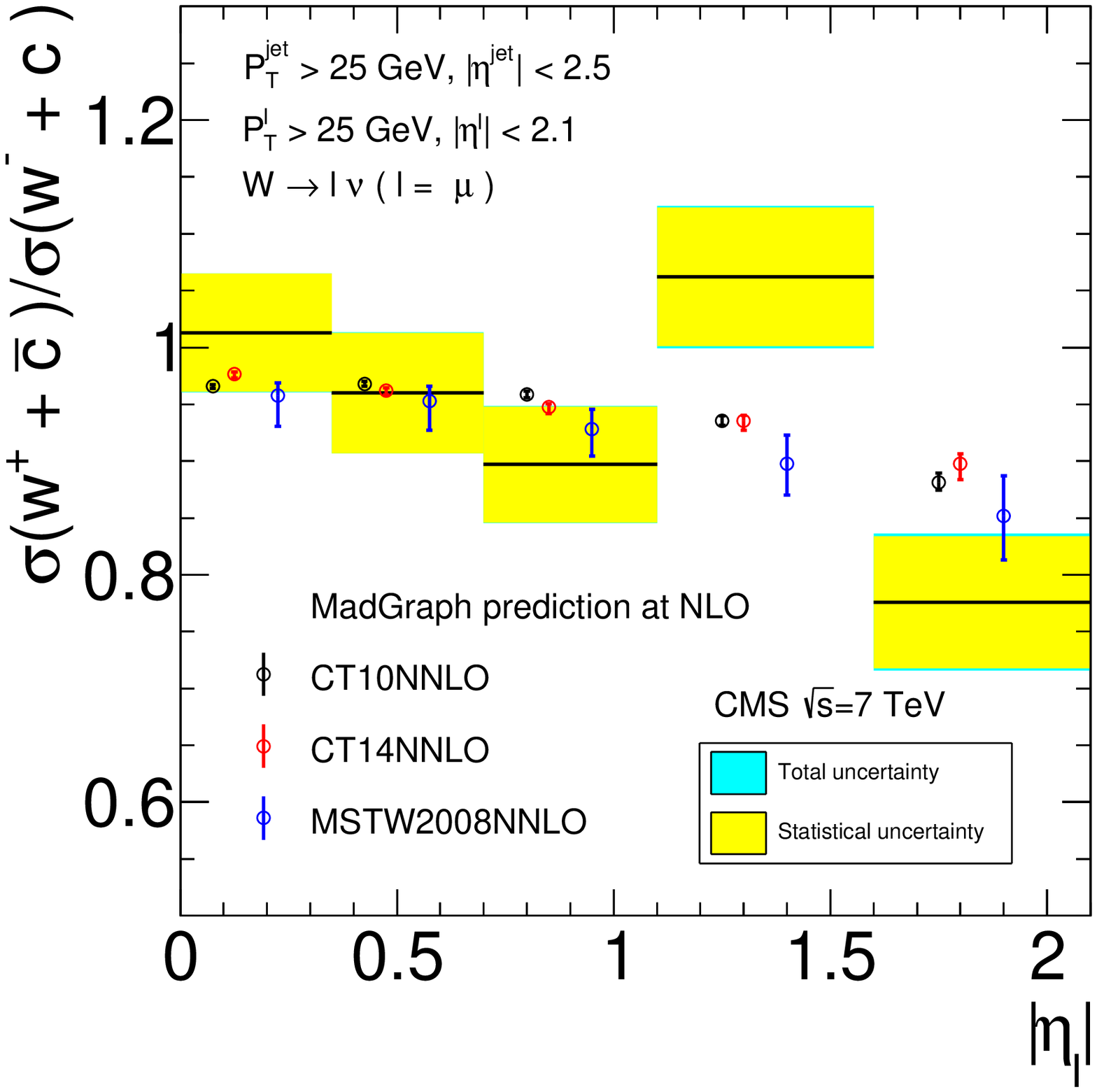}
	\includegraphics[width=0.45\textwidth]{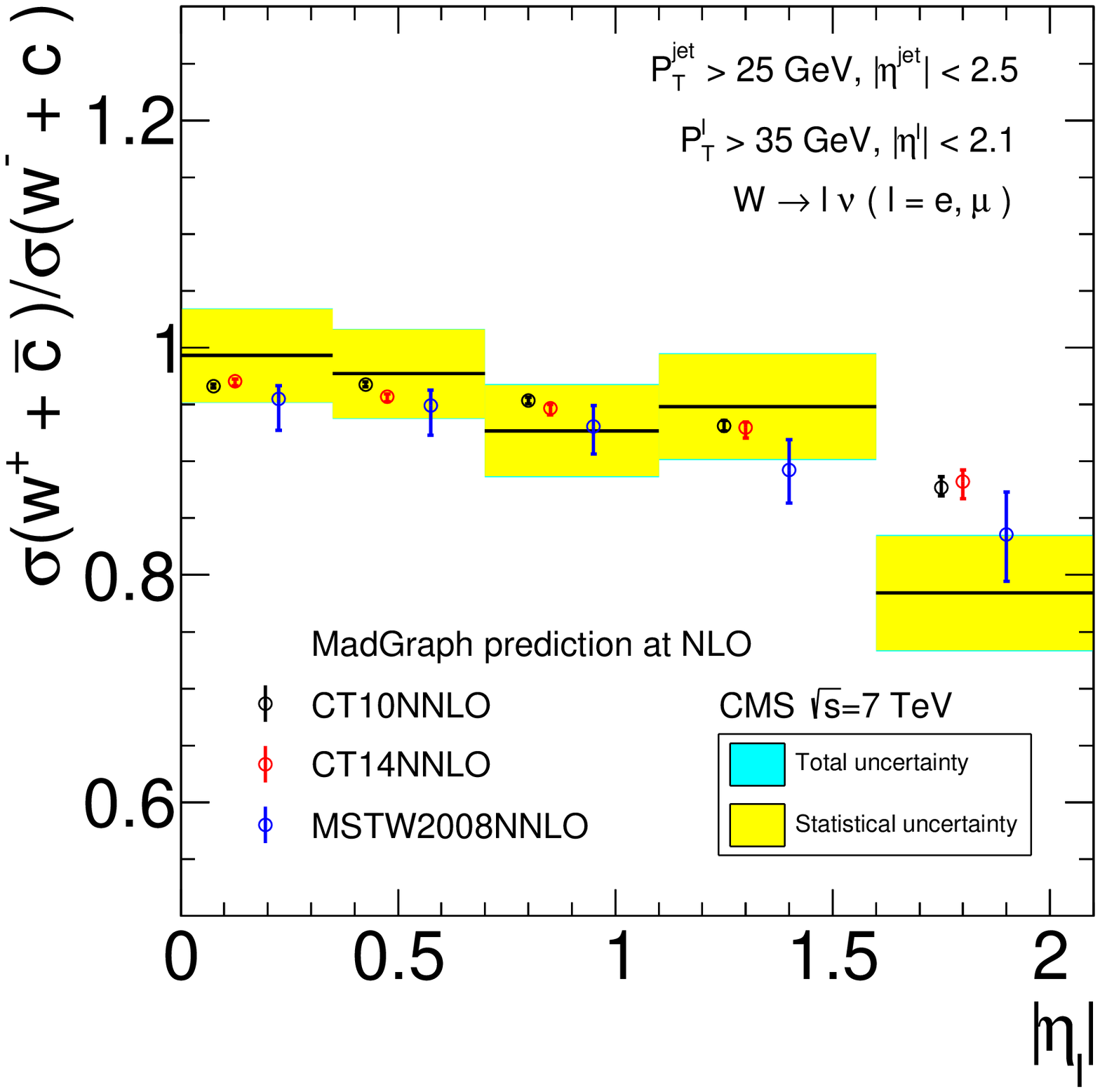}
\caption{Comparison of theoretical prediction of differential cross-section ratio for three different PDF sets  with CMS data.
}
\label{diffratioxsecfig}
\end{center}
\end{figure}

We calculated total ($\sigma(W^- + c), \sigma(W^+ + \bar c)$) and differential (absolute and normalized) cross-sections independently
under the same conditions in Subsections \ref{tXsec} and \ref{diffXsec}.
The CMS\cite{CMS7} collaboration  introduced the charged cross-section ratio,
\begin{eqnarray}\label{eq:rcpm}
R^c=\frac{\sigma(W^+ + \bar{c})}{\sigma(W^- + c)}.
\end{eqnarray}
The advantage of using this ratio is that many of the theoretical and experimental uncertainties can cancel.
The comparison of the total cross-section ratio and differential cross-section ratio with PDF uncertainty at 68\% C.L. with CMS data are shown in Fig.\ref{totalratiofig} and Fig.\ref{diffratioxsecfig}, the left column corresponds to $p_T^l > 25$ GeV, and right one is for $p_T^l > 35$ GeV.
The total cross-section ratio and differential cross-section ratio are also summarized in Table \ref{totalratiotab} and Table \ref{diffratiotab}.
From tables \ref{totalratiotab} and \ref{diffratiotab}, we see that
the total cross-section ratio, differential cross-section ratio, and the associated PDF uncertainties  are different for the CT10, CT14, and MSTW2008 PDF sets.
These differences arise from the parametrization assumptions in each global analysis. For example, the CT10 and CT14 PDF sets assume $s(x,Q) = \bar s(x,Q)$, cross-section ratios  almost exclusively are determined by the $d-\bar d$ asymmetry and with a very small PDF uncertainty.
In contrast, the MSTW08 PDF set assumes asymmetric strangeness
$s(x,Q) - \bar s(x,Q) \neq 0$, that yields a larger PDF uncertainty in the prediction.

\begin{table}[H]
	\centering
	\caption{Theoretical prediction of total cross-section ratio of $\sigma(W^+ + \bar c)/\sigma(W^- + c)$ }
	\label{totalratiotab}
	\begin{tabular}{ c |c | c }
		\hline\hline
		&\multicolumn{2}{c}{$\sigma(W^+ + \bar c)/\sigma(W^- + c)$}\\
		PDF sets&{$p^l_T>25$GeV} &{$p^l_T>35$GeV}\\
		\hline
		CT10NNLO & 0.944 $^{+0.3\%}_{-0.3\%} $& 0.942 $^{+0.4\%}_{-0.3\%} $\\
		CT14NNLO & 0.946 $^{+0.4\%}_{-0.7\%} $ & 0.940 $^{+0.4\%}_{-0.7\%} $\\
		MSTW2008NNLO & 0.920  $^{+2.2\%}_{-2.7\%} $ & 0.916 $^{+2.3\%}_{-2.7\%} $\\
		\hline
		\hline
		CMS  & {0.954 $\pm$ 2.5\% (stat.) $\pm$ 0.4\% (syst.)} &{ 0.938 $\pm$ 2.0\% (stat.) $\pm$ 0.6\% (syst.)} \\
		\hline
	\end{tabular}
\end{table}

\begin{table}[H]
	\centering
	\caption{Theoretical prediction of differential cross-section ratio with PDF uncertainty at 68 \% C.L., last column depicts CMS 7TeV measurement with statistical and systematic uncertainty.}
	\label{diffratiotab}
	\begin{tabular}{ c | c c c | c }
		\hline\hline
		\multicolumn{5}{c}{$p_T^l>25$GeV}\\
		$ |\eta^l|$ & CT10NNLO & CT14NNLO & MSTW2008NNLO& CMS measurement\\
		\hline
		$ [0,0.35]  $ &$0.966^{+0.2\%}_{-0.2\%}$&$0.977^{+0.2\%}_{-0.4\%}$&$0.958^{+1.2\%}_{-2.8\%}$&$1.013\pm5.1\%\pm0.5\%$\\
		$ [0.35,0.7]$ &$0.968^{+0.2\%}_{-0.2\%}$&$0.962^{+0.2\%}_{-0.4\%}$&$0.953^{+1.4\%}_{-2.7\%}$&$0.960\pm5.5\%\pm0.5\%$\\
		$ [0.7,1.1] $ &$0.959^{+0.3\%}_{-0.3\%}$&$0.947^{+0.3\%}_{-0.6\%}$&$0.928^{+1.9\%}_{-2.6\%}$&$0.897\pm5.7\%\pm0.9\%$\\
		$ [1.1,1.6] $ &$0.935^{+0.5\%}_{-0.4\%}$&$0.935^{+0.5\%}_{-0.9\%}$&$0.898^{+2.8\%}_{-3.1\%}$&$1.062\pm5.7\%\pm1.3\%$\\
		$ [1.6,2.1] $ &$0.881^{+1.0\%}_{-0.8\%}$&$0.898^{+1.0\%}_{-1.5\%}$&$0.852^{+4.1\%}_{-4.6\%}$&$0.776\pm7.5\%\pm2.1\%$\\
		\hline\hline
		\multicolumn{5}{c}{$p_T^l>35$GeV}\\
		$ |\eta^l|$ & CT10NNLO & CT14NNLO & MSTW2008NNLO& CMS measurement\\
		\hline
		$ [0,0.35]  $ &$0.966^{+0.2\%}_{-0.2\%}$&$0.970^{+0.2\%}_{-0.4\%}$&$0.955^{+1.2\%}_{-2.9\%}$&$0.993\pm4.1\%\pm0.7\%$\\
		$ [0.35,0.7]$ &$0.968^{+0.2\%}_{-0.2\%}$&$0.957^{+0.3\%}_{-0.5\%}$&$0.949^{+1.4\%}_{-2.7\%}$&$0.977\pm4.0\%\pm0.7\%$\\
		$ [0.7,1.1] $ &$0.953^{+0.3\%}_{-0.3\%}$&$0.947^{+0.3\%}_{-0.6\%}$&$0.931^{+2.0\%}_{-2.6\%}$&$0.927\pm4.3\%\pm0.9\%$\\
		$ [1.1,1.6] $ &$0.931^{+0.5\%}_{-0.5\%}$&$0.930^{+0.6\%}_{-1.0\%}$&$0.892^{+3.0\%}_{-3.3\%}$&$0.948\pm4.9\%\pm1.1\%$\\
		$ [1.6,2.1] $ &$0.877^{+1.1\%}_{-0.8\%}$&$0.882^{+1.2\%}_{-1.7\%}$&$0.836^{+4.5\%}_{-5.0\%}$&$0.784\pm6.4\%\pm1.4\%$\\
		\hline
		\hline
	\end{tabular}
\end{table}

\subsection{Correlation  between the CT14NNLO  and predictions for $W+$ charm data}\label{sec:cor}

 One way to determine the sensitivity of a specific data point to some PDF $f_i(x, Q)$ at a given $x$ and $Q$ is to compute a correlation cosine between the theoretical prediction for this point and the PDFs of various flavors \cite{Nadolsky:2001yg,Pumplin:2001ct,Nadolsky:2008zw}.
 Therefore, we will study the correlations between CT14NNLO  PDFs of various flavors at specific $x$ and each data point of CMS 7TeV  $W+$ charm production  with transverse momentum of the charged lepton from $W$ boson decay at the $p_T^l>35$ GeV region. However first we briefly provide the definition of the correlation cosine.
 If there are two variables $X(\vec{a_j})$ and $Y(\vec{a_j})$ in the parameter space, where ${a_j}$ are the PDF parameters, then the correlation cosine can be expressed as:
 \begin{eqnarray} \label{eq2}
 cos\phi=\frac{\vec\nabla X \cdot \vec\nabla Y}{|\vec\nabla X| |\vec\nabla Y|},\nonumber
 \end{eqnarray}
 where $\vec\nabla X$ and $\vec\nabla Y$ are gradient of the variables $X(a_j)$ and $Y(a_j)$.  For $X(a_i)$, $i$th component of gradient vector is
  \begin{eqnarray} \label{eq3}
 (\vec\nabla X)_j = \frac{\partial X}{\partial a_j}=\frac{1}{2}(X^+(a_j)-X^-(a_j)),\nonumber
 \end{eqnarray}
 where $X^+(a_j)$ and $X^-(a_j)$ are computed from the two  sets of PDFs along the positive and negative direction  of the $i$-th eigenvector. The quantity $cos\phi$ characterizes whether the variables $X$ and $Y$ are correlated ($cos\phi\sim 1$), anti-correlated ($cos\phi\sim -1$) or not correlated ($cos\phi\sim0$).

\begin{figure}[H]
\begin{center}
    \includegraphics[width=0.45\textwidth]{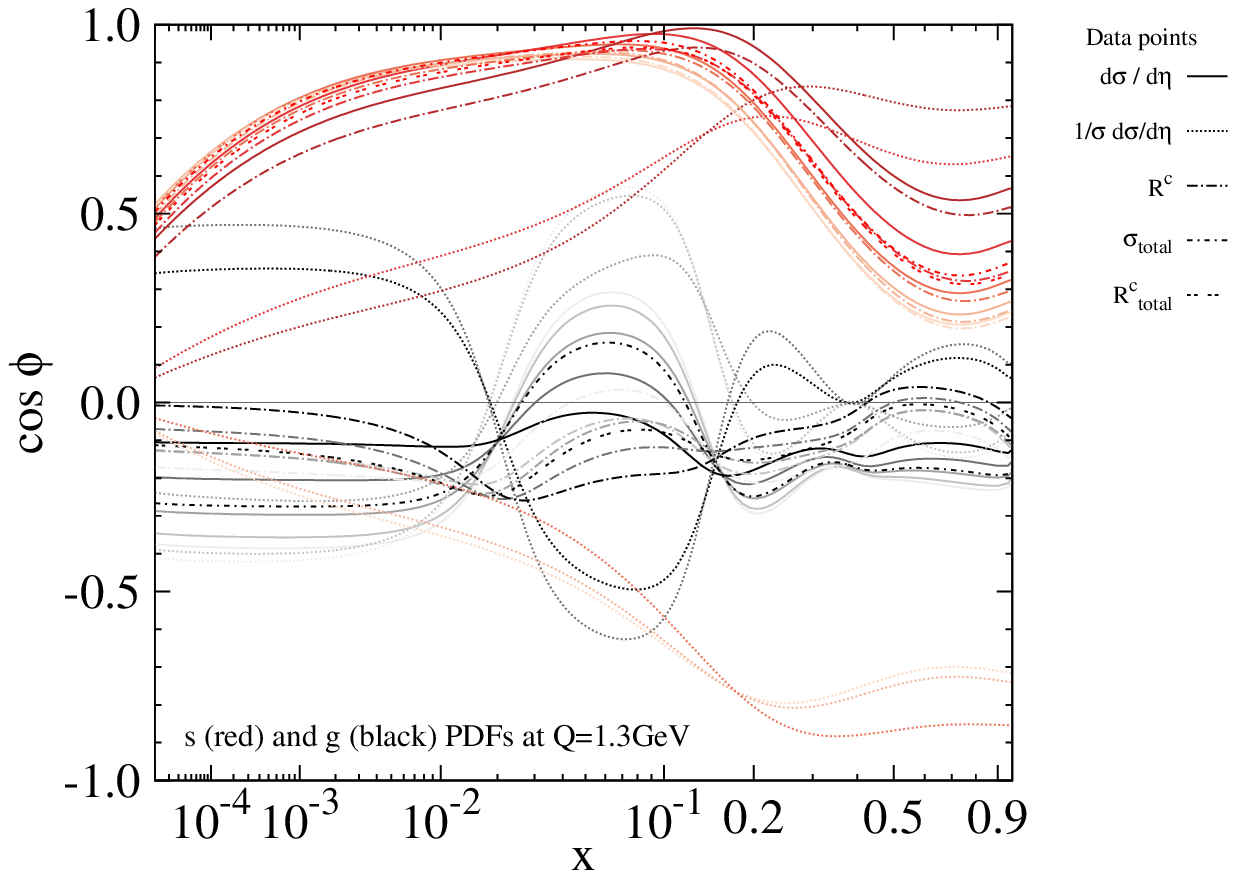}
    \includegraphics[width=0.45\textwidth]{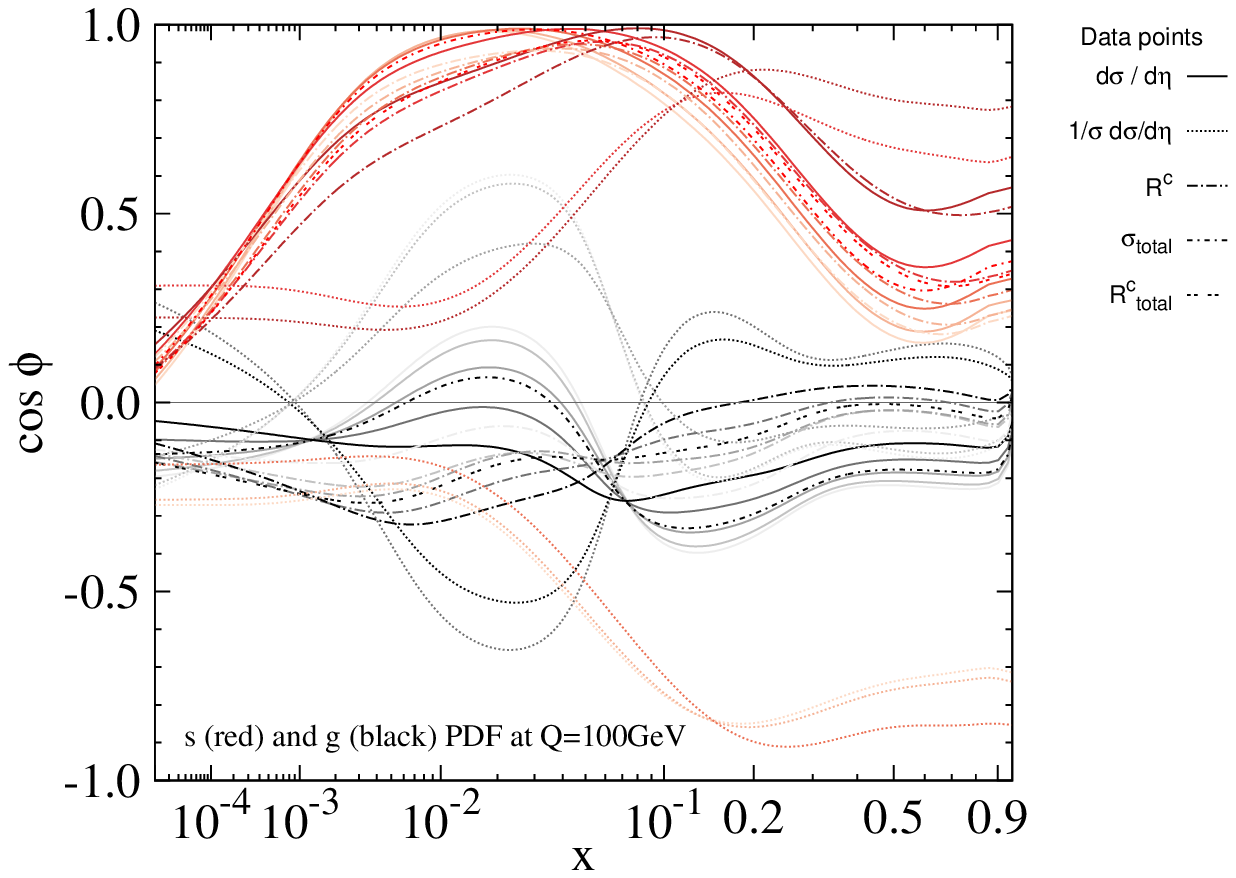}
    \includegraphics[width=0.45\textwidth]{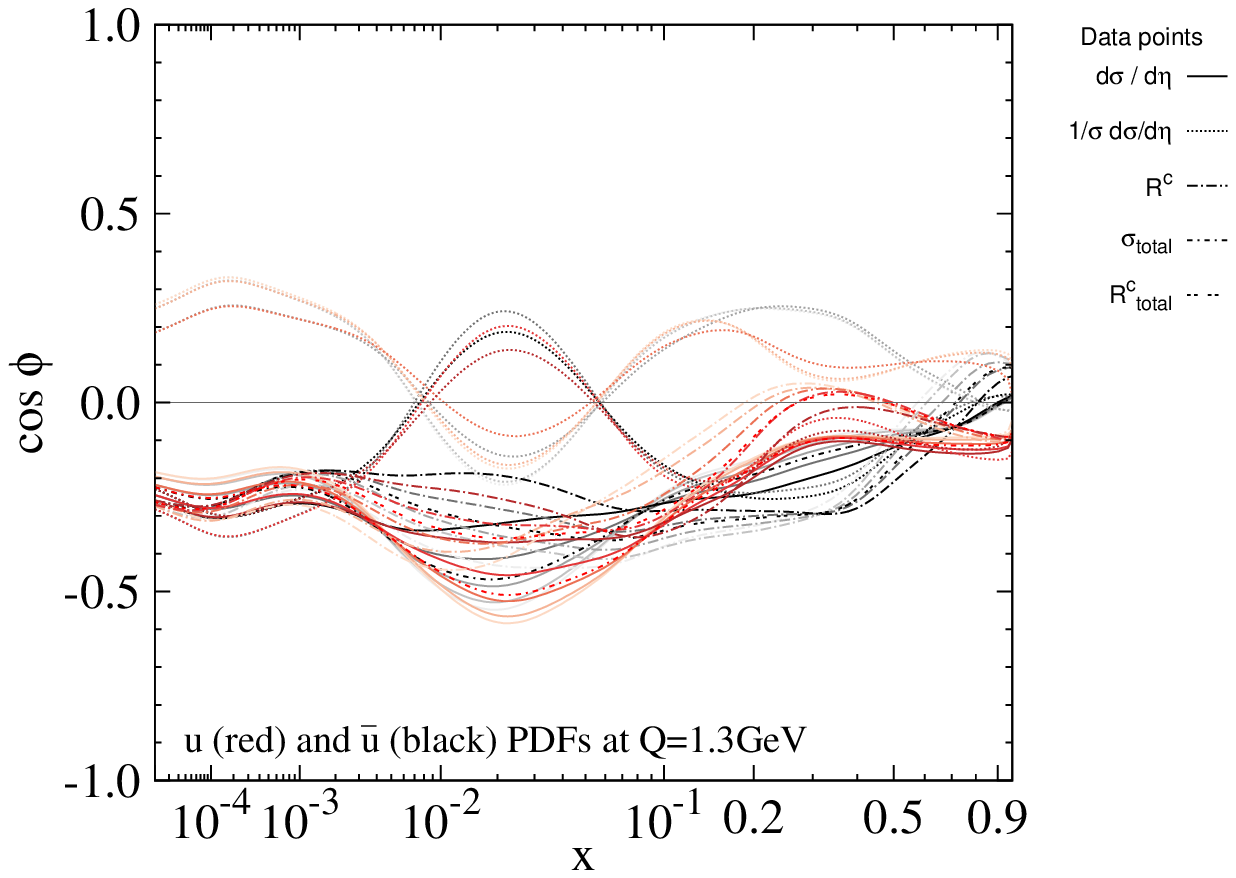}
    \includegraphics[width=0.45\textwidth]{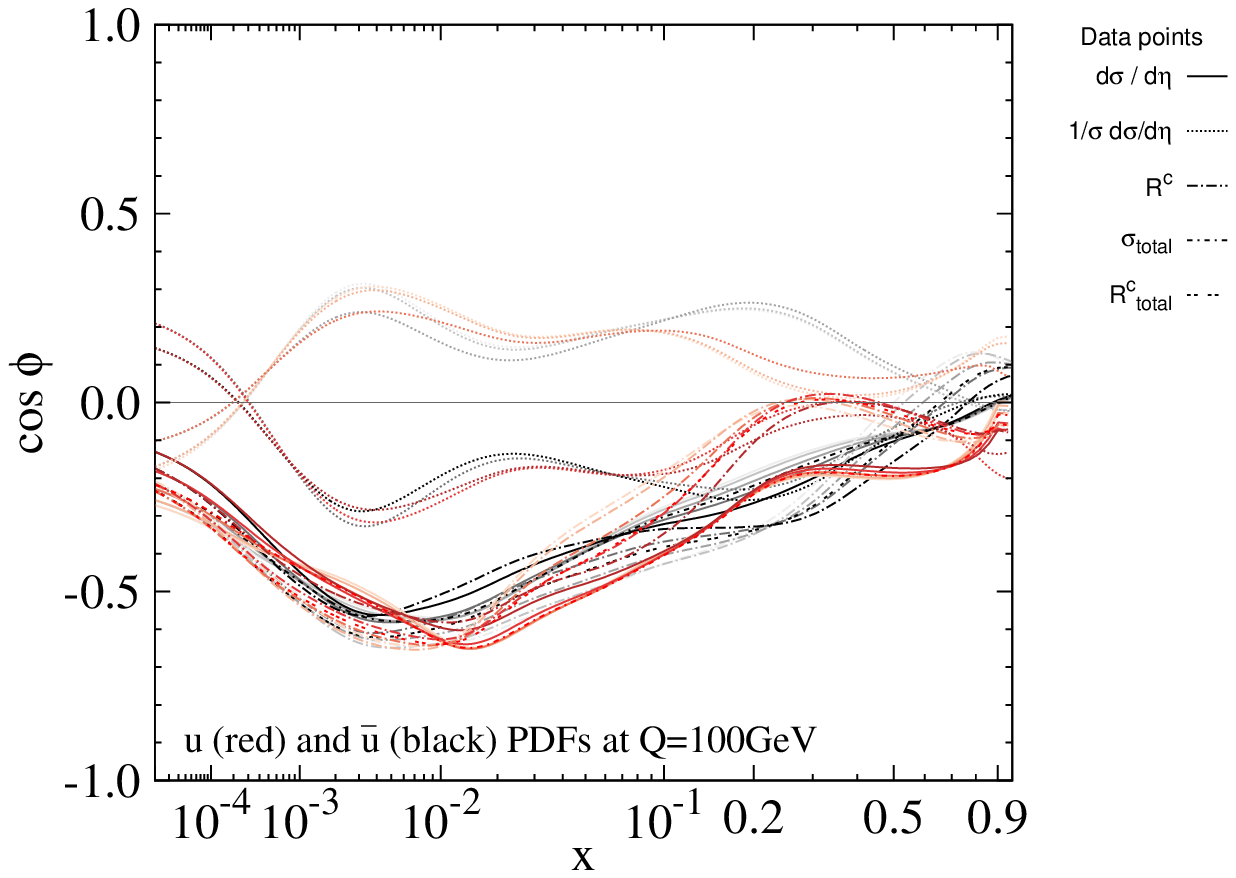}
    \includegraphics[width=0.45\textwidth]{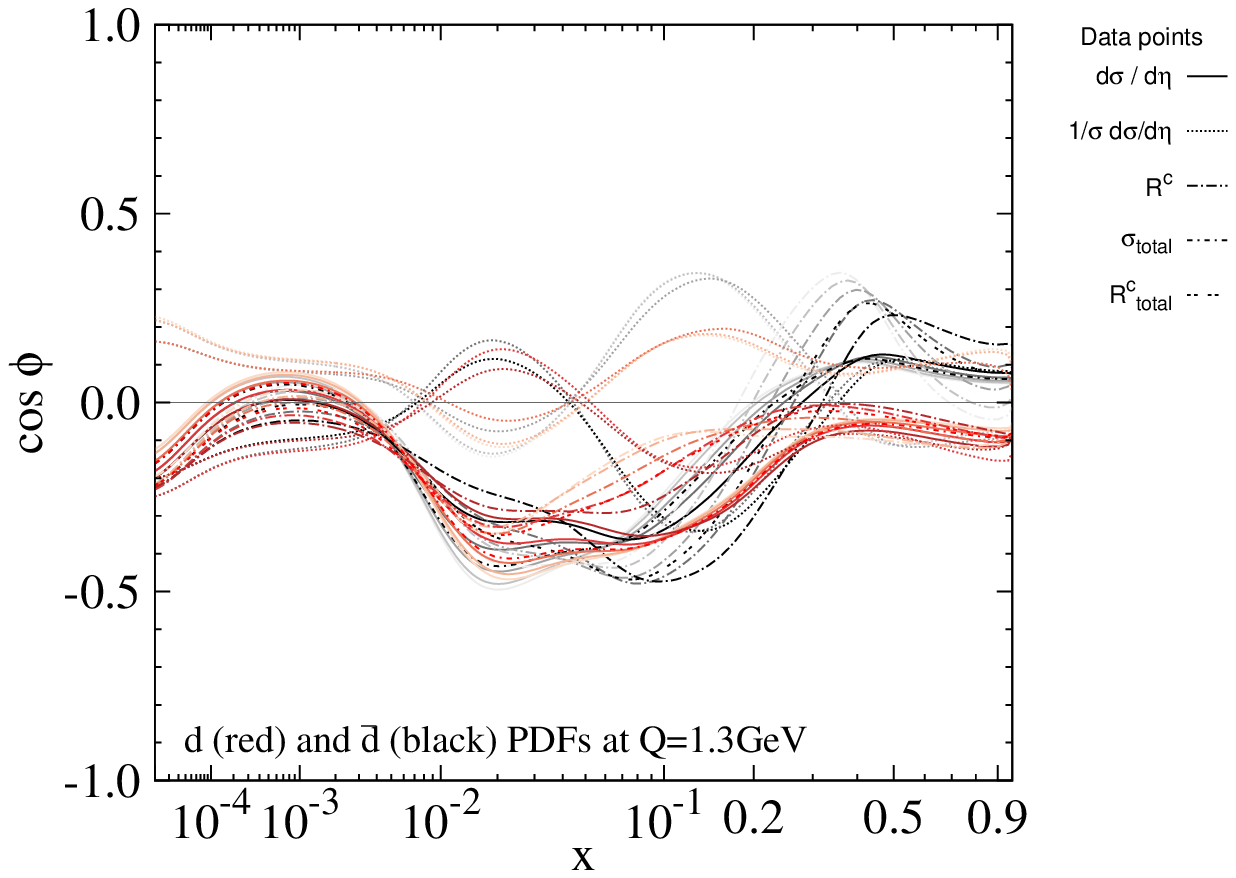}
    \includegraphics[width=0.45\textwidth]{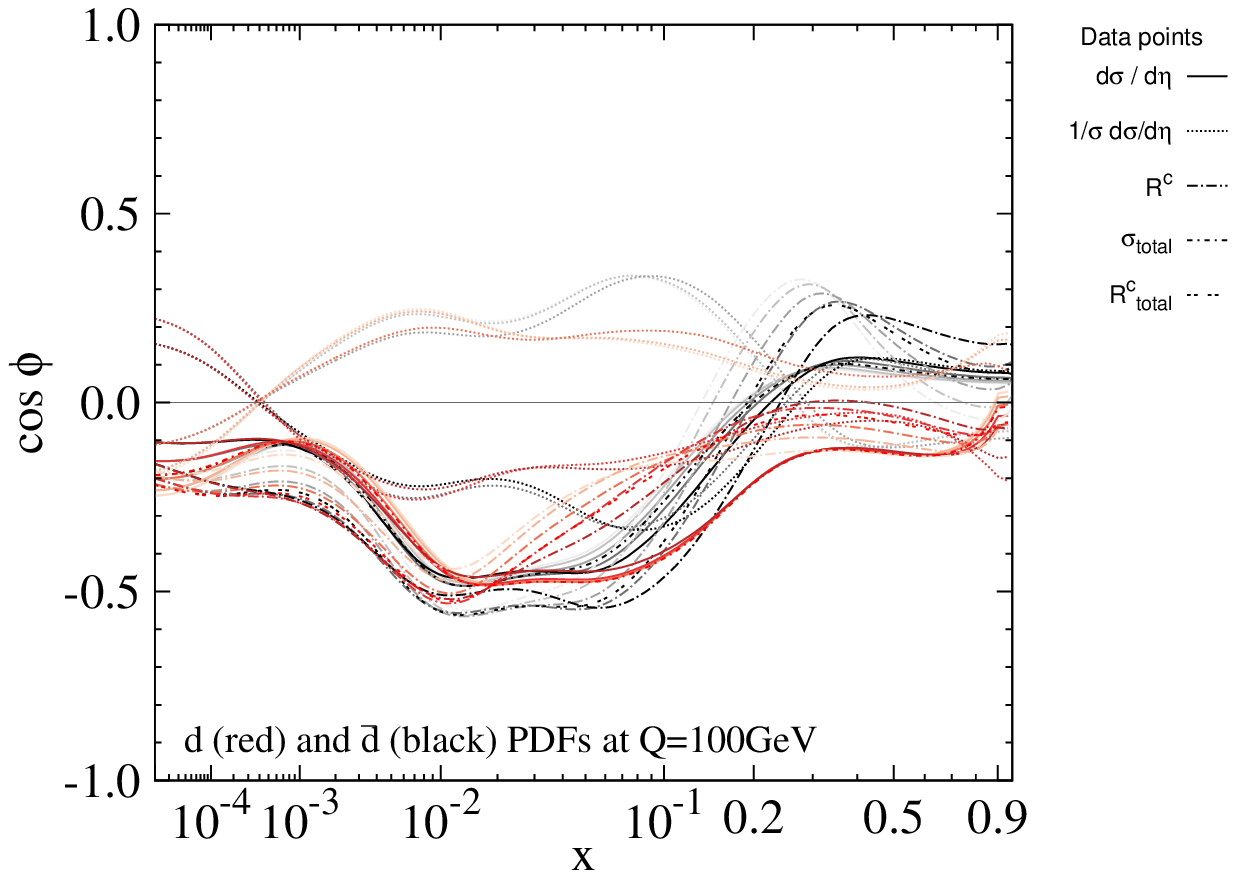}
\caption{Correlation $\cos\phi$ between CT14NNO PDFs at the specific $x$ value on the horizontal axis and CT14NNLO predictions at NLO for CMS 7 TeV $W+$ charm production at $Q=1.3$ GeV(left panel) and $Q=100$ GeV(right panel). }
\label{pdf:cor}
\end{center}
\end{figure}

 Fig.\ref{pdf:cor} shows the correlation cosine between each data point  and CT14NNLO PDF flavors at $Q=1.3$ GeV and $Q=100$ GeV. CMS 7TeV $W+$ charm data contains 17 data points, and thus there are 17 lines for certain flavors in Fig.\ref{pdf:cor}. Correlations of $s(x,Q)$ PDF and $g(x,Q)$ PDF with each data point are given at the first row in Fig.\ref{pdf:cor}.  Correlations of $u(x,Q)$ PDF and $\bar u(x,Q)$ PDF with each data point are given at the second row in Fig.\ref{pdf:cor}. Finally correlations of $d(x,Q)$ PDF and $\bar d(x,Q)$ PDF with each data point are given at the third row in Fig.\ref{pdf:cor}. In each subfigure, the correlation between one of the PDF flavors  with each data point is distinguished by different type of line. Solid, long-dashed-dotted, dotted, short-dashed, and short-dashed-dotted lines correspond to correlation of differential cross-section, differential cross-section ratio, normalized differential cross-section, total cross-section, and total cross-section ratio data respectively. As we discussed in Section \ref{sec3}, differential cross-section, differential cross-section ratio and normalized differential cross-section data has included five data points that are measured by five rapidity bin ranges. The lines with darker color correspond to higher rapidity bin range.
In the case of the total cross-section, differential cross-section and ratio, $s(x,Q)$ PDF correlations are most significant ($\cos{\phi}\sim 1$) at $x$ from few times $10^{-2}$ to few times $10^{-1}$, when $Q=1.3$ GeV and $Q=100$ GeV respectively. However at other $x$ range, the $s(x,Q)$ PDFs correlation is not very strong. There are no clear relations between the rapidity bin range and correlation cosine, however it can be seen from two the subfigures at the first row that each data point in various rapidity bins has a strong correlation with $s(x,Q)$ PDF at the $x$-region mentioned above. There is other information for the normalized differential cross-section, which includes five data points that are partially correlated and partially anti-correlated, and represented with each flavor, as illustrated in each subfigure of Fig.\ref{pdf:cor} with five dashed-lines (red or blue) that are inconsistent with other types of lines. Correlated data points prefer PDFs to become larger and anti-correlated data points prefer to PDFs to become smaller. Hence the impact of normalized cross-section data on PDFs is neutralized. Not considering the correlations with normalized cross-section data, the $d$(anti) quark and $u$(anti) PDFs are anti-correlated at $x<0.4$ region, and small at all $x$-regions when both $Q=1.3$ GeV and $Q=100$ GeV, which can be seen in subfigures in the second and third rows of Fig.\ref{pdf:cor}. Thus those data points without normalized cross-section data prefer $d$(anti) quark and $u$(anti) PDFs to become smaller in the $x<0.4$ region. Gluon's correlation is likewise miniscule in the $x$-region which can be seen in the subfigures in the first row of Fig.\ref{pdf:cor}. One can conclude that CMS 7TeV $W+$ charm data have larger impact on $s(x,Q)$ PDF in CT14NNLO than other flavors.

 \section{Using EPUMP to study the impact of $W+$ charm data on CT14NNLO PDFs}\label{sec4}

\begin{figure}[H]
\begin{center}
    \includegraphics[width=0.45\textwidth]{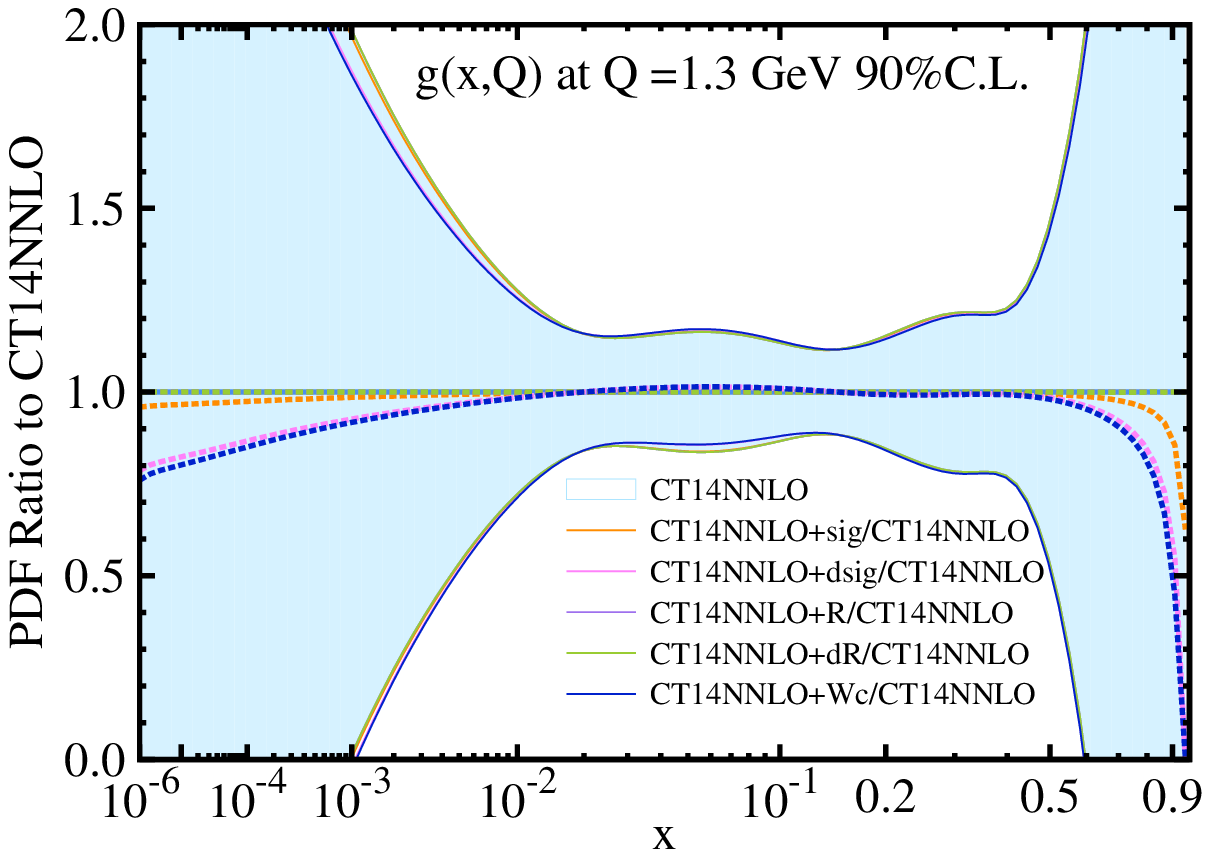}
    \includegraphics[width=0.45\textwidth]{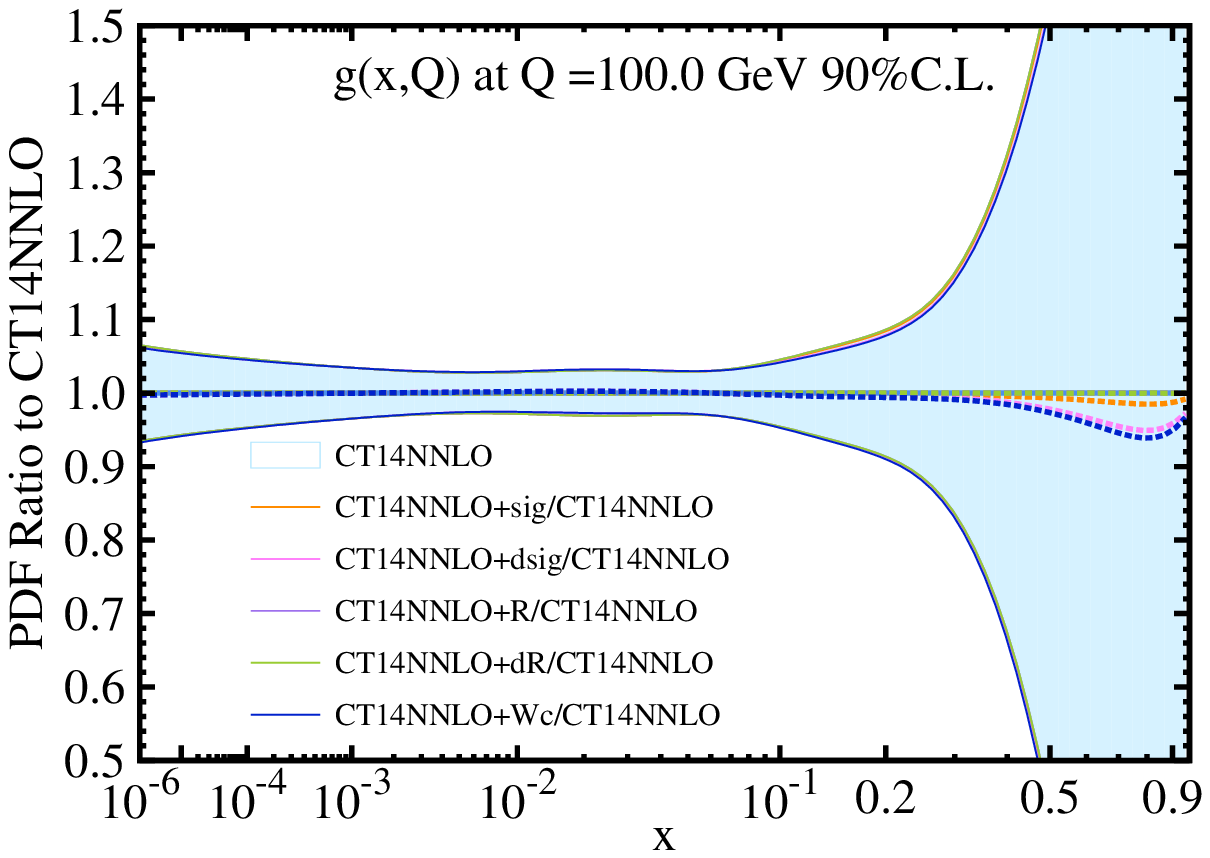}
    \includegraphics[width=0.45\textwidth]{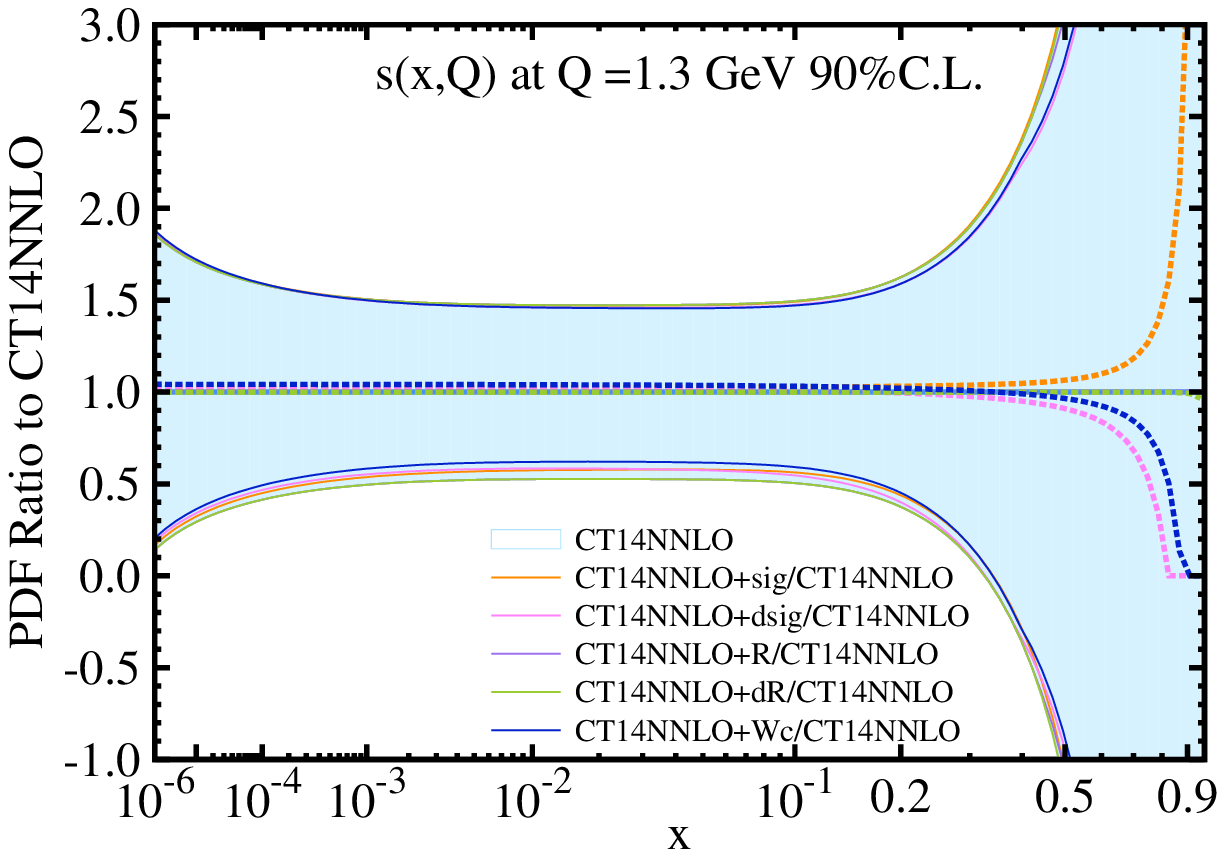}
    \includegraphics[width=0.45\textwidth]{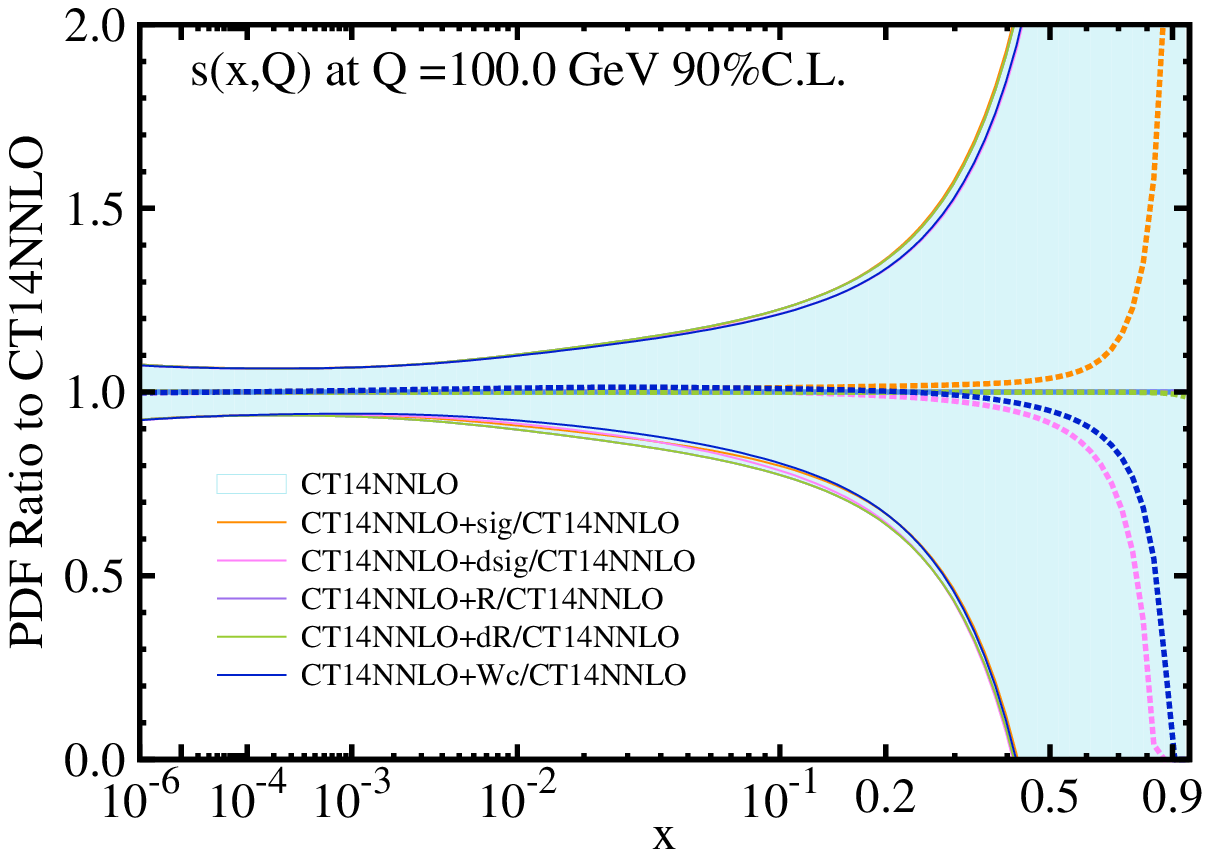}
    \caption{ Comparison of 90\% C.L. $g(x,Q)$ PDF (first row)  and $s(x,Q)$ PDF (second row) uncertainties from CT14NNLO and  CT14NNLO+sig, CT14NNLO+dsig, CT14NNLO+R, CT14NNLO+dR, CT14NNLO+Wc.
   Shaded area represents error bands of  CT14NNLO PDFs. The area between solid line represents error bands of ePump-updated  PDFs, which  are distinguished by different colors, and dotted line depicts the best fit PDFs. Both best fit PDFs and error bands are normalized to the corresponding central CT14NNLO PDFs.}
\label{pdf:sq1}
\end{center}
\end{figure}

\begin{figure}[H]
	\begin{center}		\includegraphics[width=0.45\textwidth]{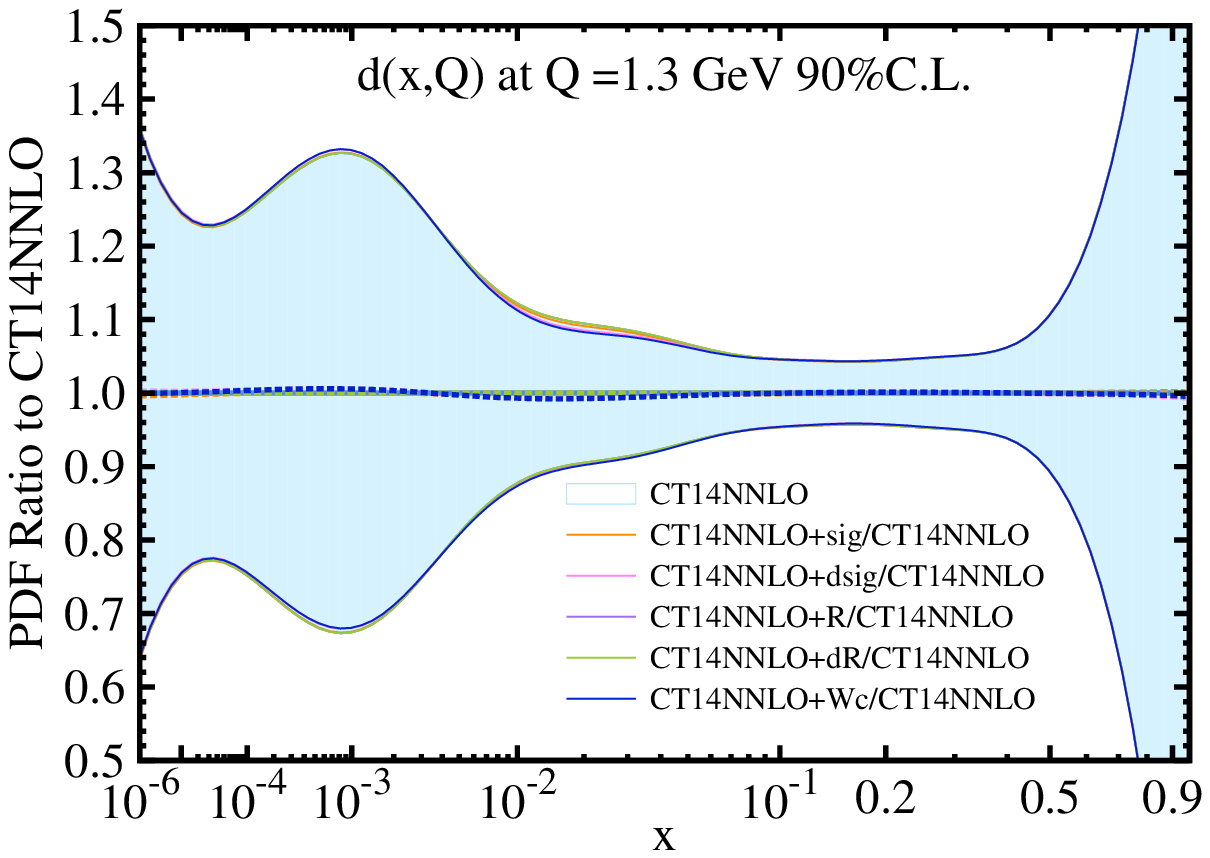}		\includegraphics[width=0.45\textwidth]{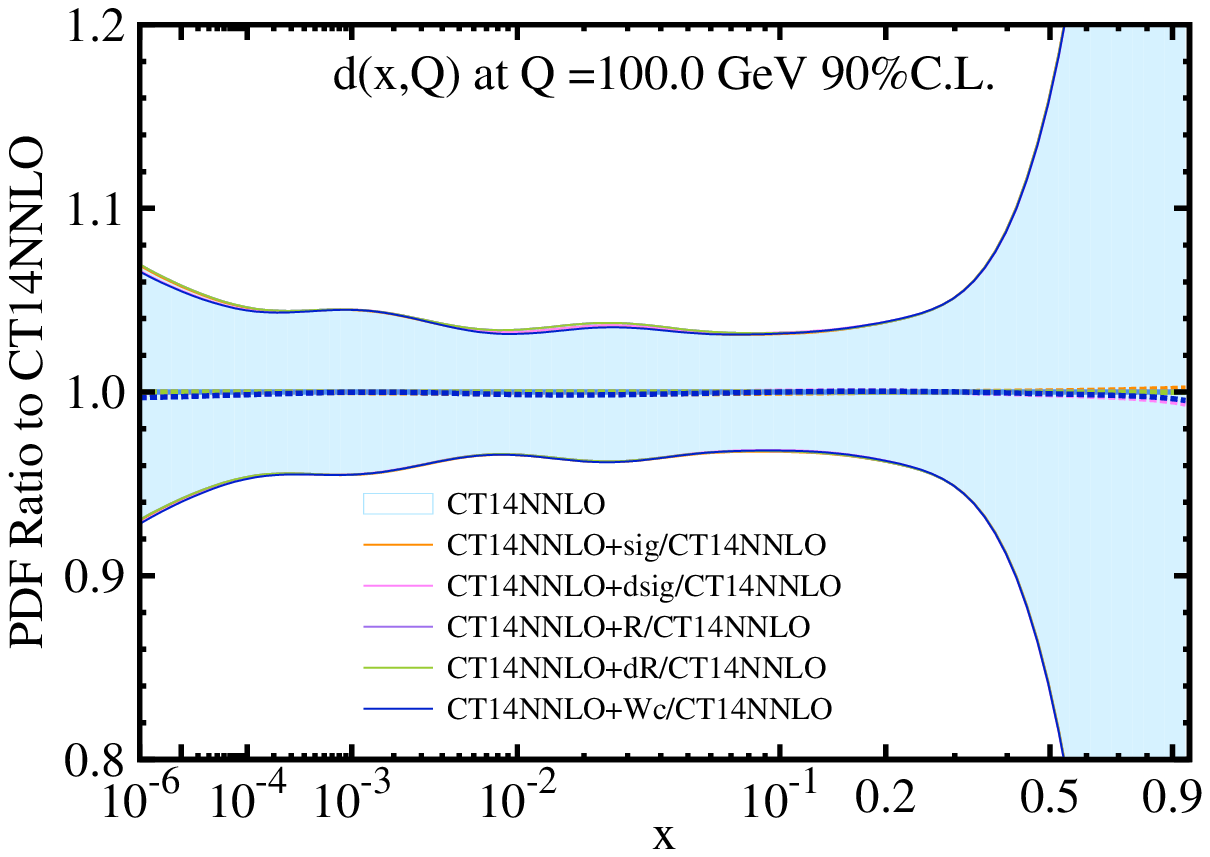}		\includegraphics[width=0.45\textwidth]{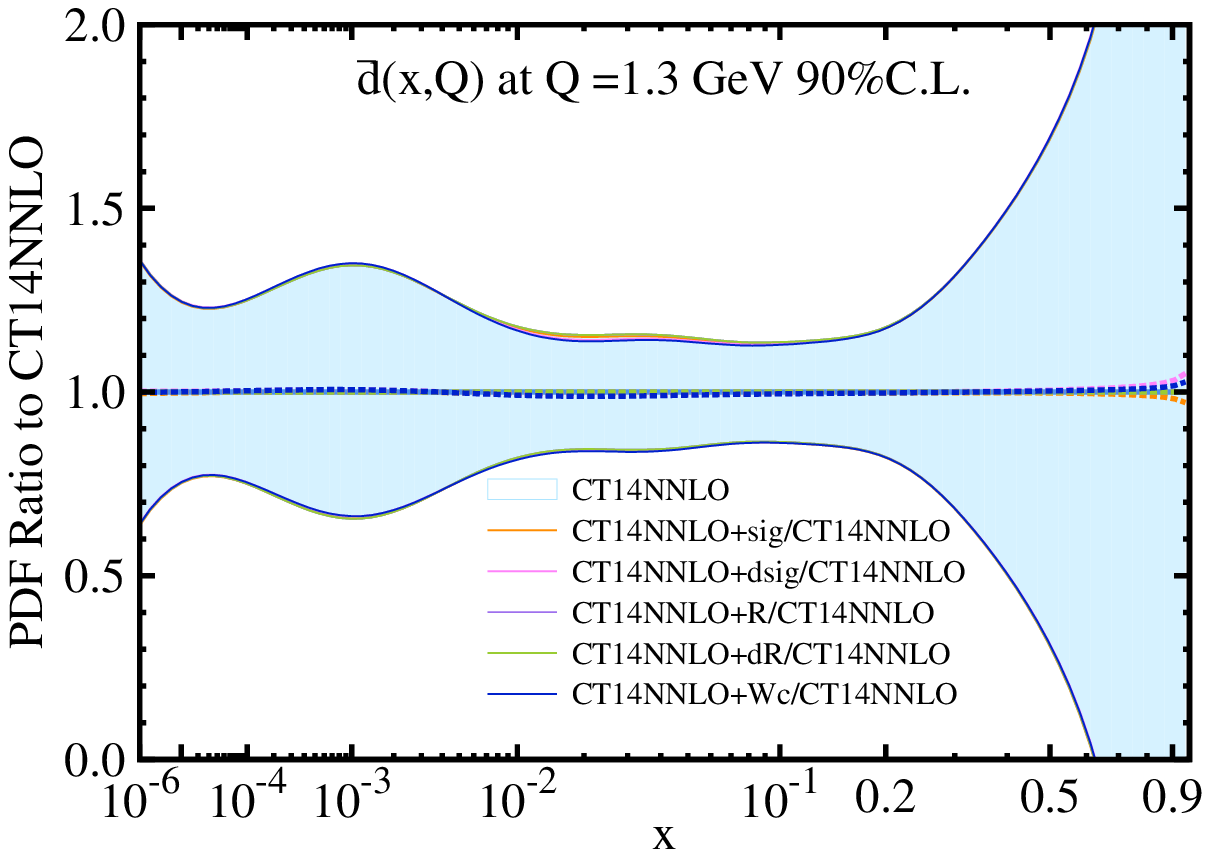}		\includegraphics[width=0.45\textwidth]{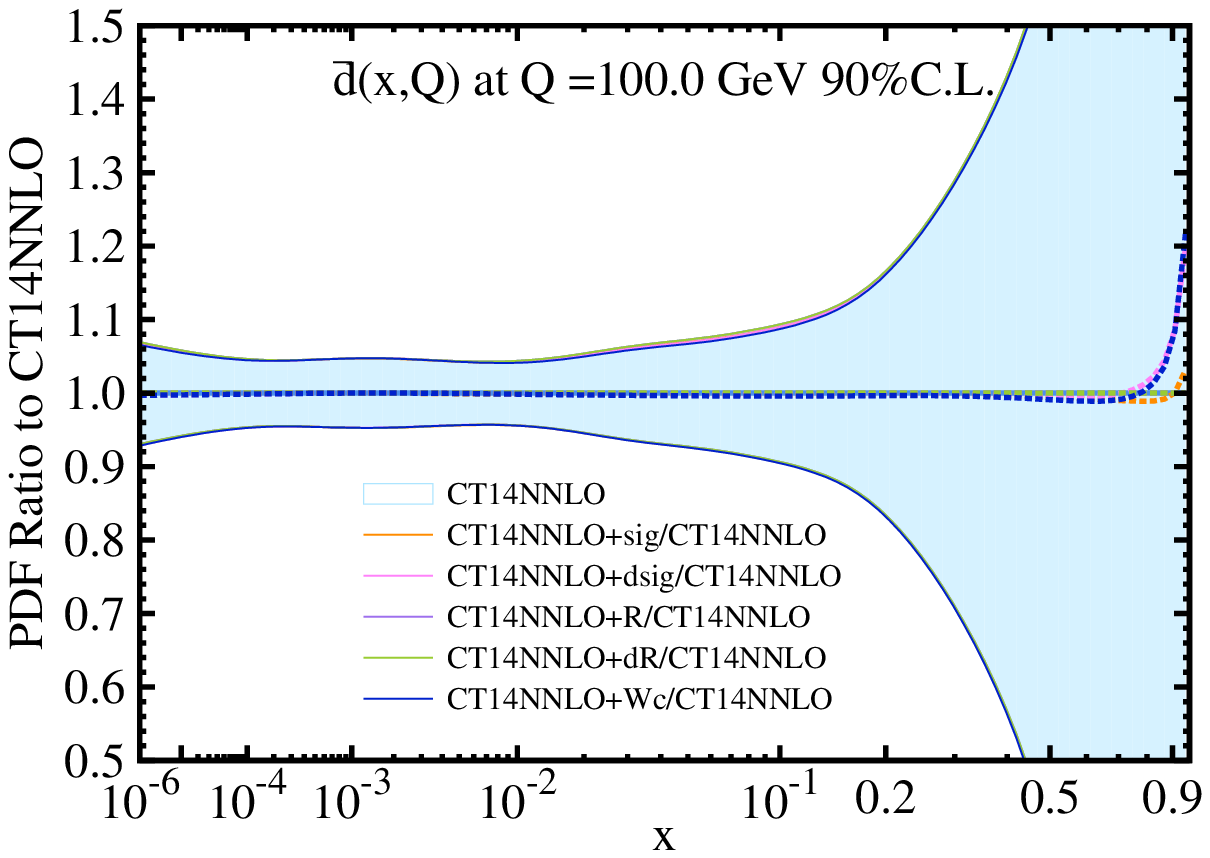}
\caption{ Comparison of 90\% C.L. $d(x,Q)$ PDF(first row)  and $\bar d(x,Q)$  PDF(second row) uncertainties from CT14NNLO and  CT14NNLO+sig, CT14NNLO+dsig, CT14NNLO+R, CT14NNLO+dR, CT14NNLO+Wc.
   Shaded area stands for  error bands of  CT14NNLO PDFs. The area between solid line stands for error bands of the epump updated  PDFs, which  are distinguished by different colors, and dotted line stands for best fit PDFs.}
		\label{pdf:sq21}
	\end{center}
\end{figure}

The study cited in Ref.\cite{Schmidt:2018hvu} presented a software package, ePump (error PDF updating method package), that can be used to update or optimize a set of PDFs, including the best-fit PDF set and error PDFs, and to update
any other set of observables. Furthermore,  Ref. \cite{Willis:2018yln} and Ref.\cite{Hou:2019gfw} cite interesting further studies using ePump.
In this section,  we use ePump to analyze the impact of CMS 7TeV $W+$ charm production measurements on the CT14NNLO PDFs.
To update CT14NNLO PDFs, we use the  CMS 7TeV  total cross-section(one data point), differential cross-section(five data points), total cross-section ratio(one data point) and differential cross-section ratio(five data points), as well as combined data sets and their NLO QCD predictions from MadGraph as ePump inputs. CT14NNLO+sig, CT14NNLO+dsig, CT14NNLO+R, CT14NNLO+dR, and CT14NNLO+Wc  in Figs.\ref{pdf:sq1}-\ref{pdf:sq21} are the ePump-updated PDFs by total cross-section data, differential cross-section data, total cross-section ratio data, differential cross-section ratio data, and combined CMS 7TeV $W+$ charm data.
The weight factor for each data is three in our ePump studies. A weight larger than one is equivalent to having more data points with the
same experimental uncertainties or, alternatively, to reducing the experiment uncertainties
by a factor of the square root of the weight.
In the combined data, we excluded the normalized differential cross-section data to avoid double counting. After updating, the relative changes in CT14NNLO ensembles are best visualized by comparing their PDF error band and PDF ratio, in which ratio plot is obtained by dividing the error set and best fit of updated PDFs by the best fit of original CT14NNLO PDFs.
In Figs.\ref{pdf:sq1}-\ref{pdf:sq21}, we show the impact of $W+c$ data and combined data on CT14NNLO PDFs, namely,
Comparisons of CT14NNLO PDF (light blue) and ePump-updated  PDFs CT14NNLO+sig (orange), CT14NNLO+dsig (magenta), CT14NNLO+R (purple), CT14NNLO+dR (green), and CT14NNLO+Wc(blue) at $Q=1.3$ GeV (left column) and $Q=100$ GeV (right column). Flavors $g(x,Q)$, $s(x,Q)$,  $u(x,Q)$, $\bar u(x,Q)$, $d(x,Q)$, and $\bar d(x,Q)$ are shown. The PDF uncertainty bands are 90\% C.L..

The change in $g(x,Q)$ PDF mostly comes from the differential cross-section data.
The changes in the best fit value and uncertainty of the ePump-updated $g(x,Q)$ PDF are visualized in the first row of Fig.\ref{pdf:sq1}, compared to CT14NNLO PDFs.
The central value of the updated  $g(x,Q)$ PDF in the range $10^{-1} < x < 0.5$ remains almost unchanged, and it is increased slightly in the range $10^{-2}<x<10^{-1}$, compared to CT14NNLO for $Q=1.3$GeV, whereas it is decreased by large factors for $x<10^{-3}$ and  $x>0.5$. For $Q=100$GeV, the central value of the updated  $g(x,Q)$ PDF in the range $ x < 0.5$ remains almost unchanged, and some variations of the best fit $g(x,Q)$ in the region $x>0.5$ are observable. The $g(x,Q)$ PDF is not efficiently determined in very small and  very large $x$-regions, however all remain within the error bands of the PDFs.
The error band of the updated $g(x,Q)$ PDF is slightly reduced in the range $10^{-2}<x<10^{-1}$ for $Q=1.3$GeV. In other regions, the updated  $g(x,Q)$ PDFs uncertainty bands are comparable to that of CT14NNLO;

The $s(x,Q)$ PDF is most sensitive to CMS 7TeV $W+$ charm data.
Total and differential cross-section data are responsible for most
of the changes in the central values and uncertainties of the updated $s(x,Q)$ PDF.
The changes in  central value and uncertainty of the ePump-updated $s(x,Q)$ PDF is visualized in the second row of Fig.\ref{pdf:sq1}, compared to CT14NNLO PDFs.
After updating the PDF by combined data, the best fit $s(x,Q)$ PDF is increased slightly in the region $x<0.4$ for $Q=1.3$GeV. 
There is a large change of the  $s(x,Q)$ PDF  in the region $x>0.5$ for both $Q=1.3$ GeV and $Q=100$ GeV, however well within the error bands of PDFs.

The LO contributions of $s$, $\bar s$, $d$, $\bar d$ quarks  to the $W+$ charm production cross-section
are shown in Table \ref{LOcon}.
  $d$ and $\bar d$ quarks contributions are significantly small relative to $s$ and $\bar s$ quarks contributions, because $dg\to W^-+c$ and $\bar d g\to W^++\bar c$ processes are suppressed by the CKM matrix element. However we can not neglect $d$ and $\bar d$ quark contributions  to the $W+$ charm cross-section, $d$  quark contribution is about 11 \%  of the $W^-+c$ productions cross-section and the $\bar d$ quark contribution is about 6 \%  of the $W^-+c$ productions cross-section. Therefor, CMS 7TeV $W+$ charm data can have an impact on both  $d(x,Q)$ and $\bar d(x,Q)$ PDFs.
 Fig. \ref{pdf:sq21} shows the changes in central values and uncertainties of the updated  $d(x,Q)$ and $\bar d(x,Q)$ PDFs  for both $Q=1.3$ GeV and $Q=100$ GeV.
 Most of the changes in $d(x,Q)$ and $\bar d(x,Q)$ PDFs come from total and differential cross-section data of CMS 7TeV $W+$ charm production. After updating the CT14NNLO PDFs by combined data, the error band of $d(x,Q)$ PDF is slightly reduced  in the region $10^{-2}<x<10^{-1}$  for  both $Q=1.3$GeV and $Q=100$GeV. The best fit $d(x,Q)$  PDF decreased a little bit in this region.
For both $Q=1.3$GeV and $Q=100$GeV, the central value and uncertainty of the ePump-updated $\bar d(x,Q)$ PDF are close to that of CT14NNLO in the region $x <0.5$;
The best fit $\bar d(x,Q)$ PDF is increased significantly in the region $x>0.5$, however well  within the error bands of PDFs.

At the LO, $u$ quark does not contribute to the $W+$ charm production cross-section, however it does so beyond the LO. In our ePump study, we employ the  theoretical prediction of the $W+$ charm production cross-section at NLO QCD. Therefore, we compare the ePump-updated PDFs via CMS 7TeV $W+$ charm data and CT14NNLO PDFs to see the impact on $u(x,Q)$ and $\bar u(x,Q)$ PDFs in CT14NNLO for both $Q=1.3$ GeV and $Q=100$ GeV.
We found that the central value and uncertainties of the ePump updated $u(x,Q)$ and $\bar u(x,Q)$ are almost unchanged.

\begin{figure}[H]
	\begin{center}
		\includegraphics[width=0.45\textwidth]{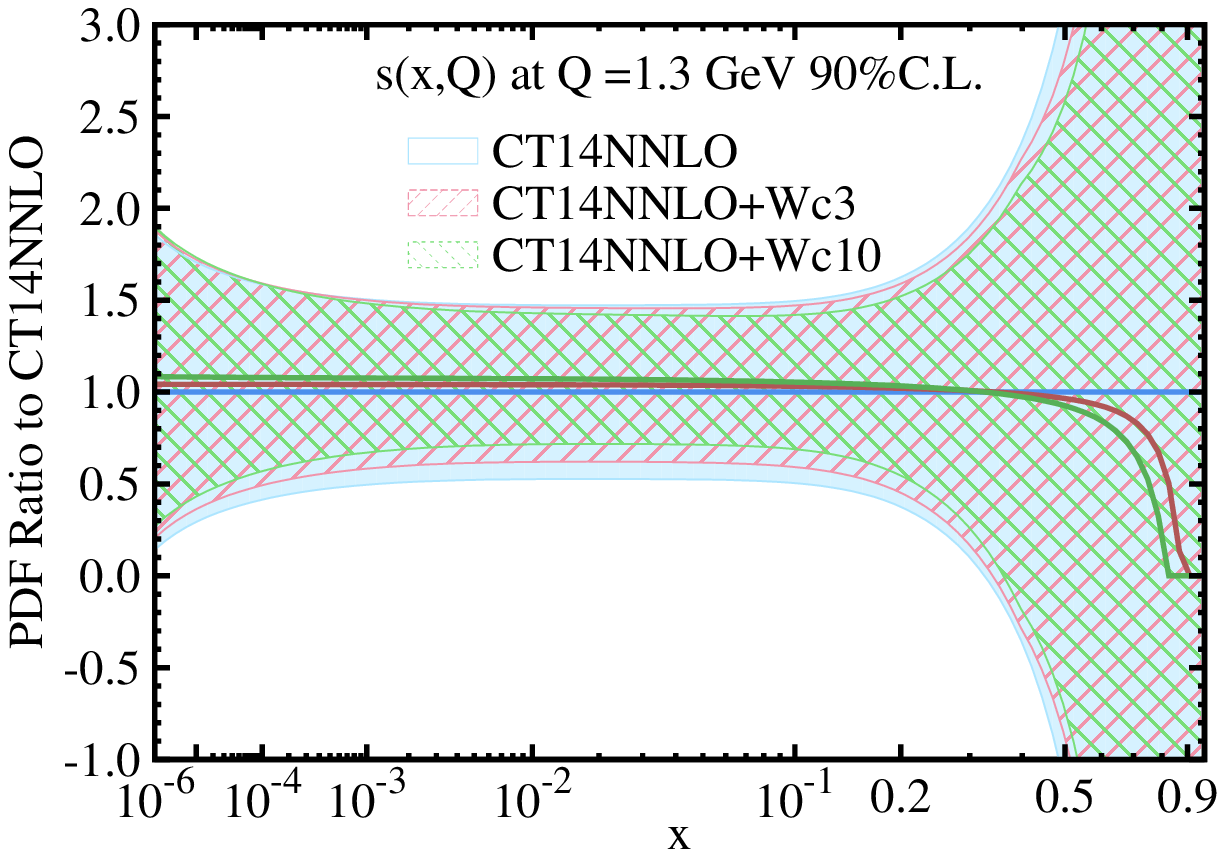}
		\includegraphics[width=0.45\textwidth]{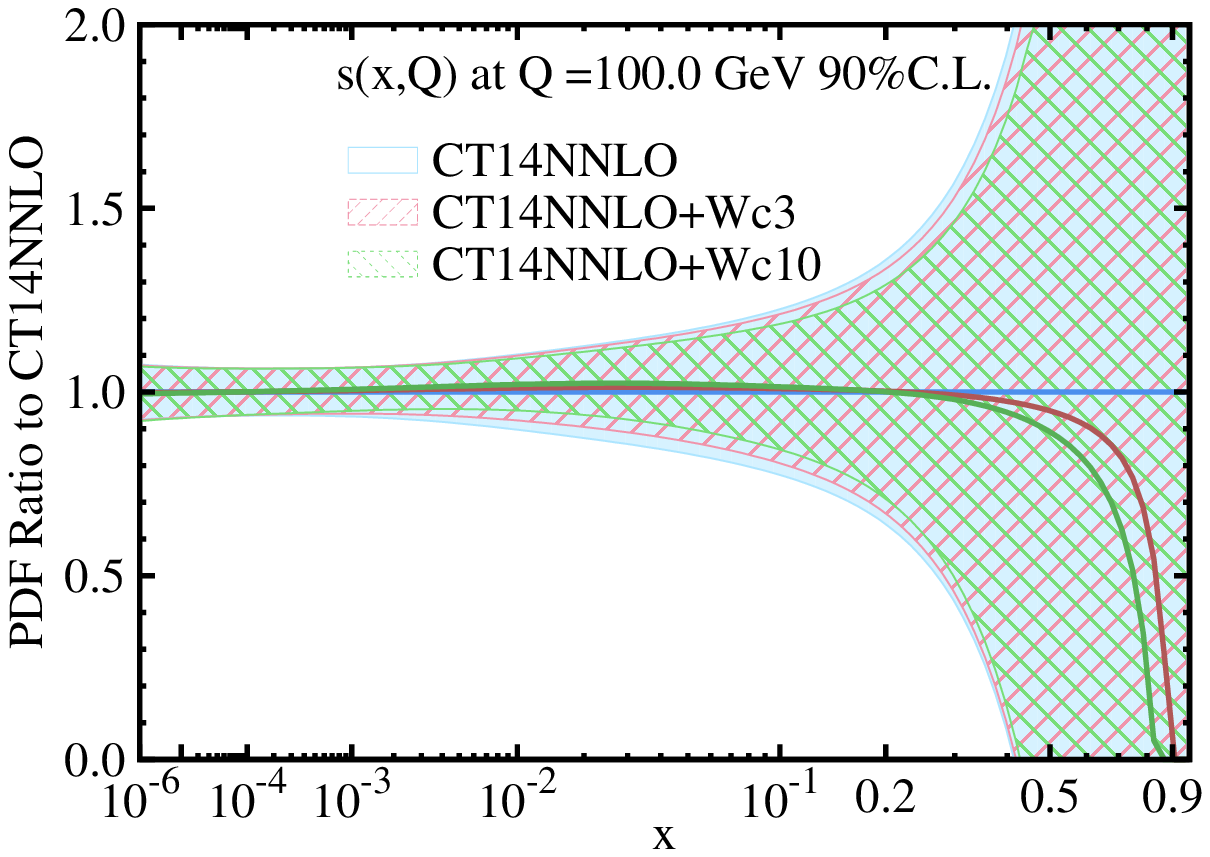}
		\caption{Comparison of 90\% C.L. $s(x,Q)$ PDF uncertainties at $Q=1.3$ GeV and $Q=100$ GeV from CT14NNLO and  CT14NNLO+Wc3 and CT14NNLO+Wc10
 corresponding to $s(x,Q)$ PDF updated by combined CMS 7TeV $W+$ charm data with weight factors three and ten. }
		\label{pdf:1to10}
	\end{center}
\end{figure}

In Fig.\ref{pdf:1to10}, we compared the $s(x, Q)$ PDF from CT14NNLO, and ePump-updated  $s(x, Q)$ PDF from combined CMS 7TeV $W+$ charm data with weights three amd ten. Fig. \ref{pdf:1to10} shows that the $s(x, Q)$ PDF error band greatly decrease at $x<0.4$ for  $Q=1.3$ GeV and $Q=100$ GeV when the weight factor is increased from three to ten. The best fit $s(x, Q)$ PDF increases in the region $ x < 0.1$. 
At large $x$, the best fit $s(x, Q)$ PDF decreases significantly
for both $Q=1.3$ GeV and $Q=100$ GeV, however remains well within the error bands of PDFs.

\begin{figure}[H]
	\begin{center}
		\includegraphics[width=0.9\textwidth]{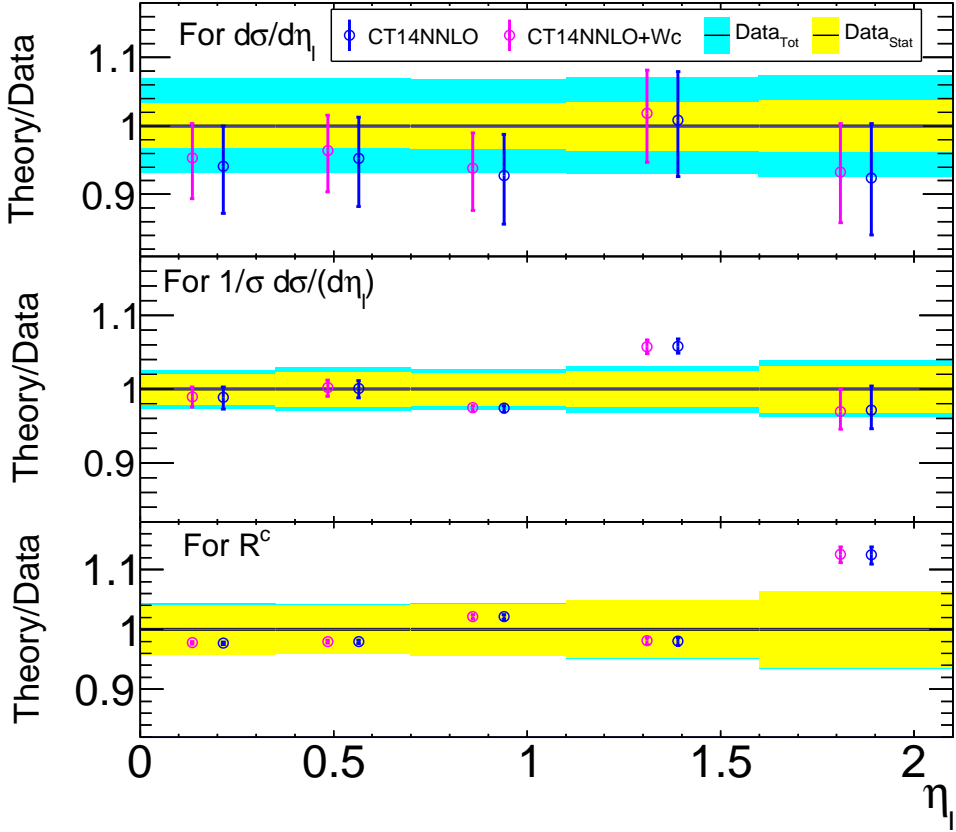}
		\caption{Comparison of data and theoretical predictions obtained from updated and original CT14NNLO PDFs, PDF uncertainty  at 68\% C.L.}
		\label{utheo}
	\end{center}
\end{figure}

Using  ePump \cite{Schmidt:2018hvu}, we updated  CT14NNLO PDFs.
One might want to know how the inclusion of the CMS 7TeV $W+$ charm  data in the global PDF fits would modify the prediction  and uncertainties for any other set of observables including the original observables that were used for updating the CT14NNLO PDFs.
ePump can also directly update predictions  and uncertainties for any  observables after including new data.
In Fig.\ref{utheo}, we compare the predictions from  CT14NNLO PDFs and  CT14NNLO+Wc PDFs, obtained by updating the CT14NNLO with  CMS 7TeV $W+$ charm data using ePump with the CMS 7TeV $W+$ charm data. Fig.\ref{utheo} shows that the uncertainties decreased after the update, and the predicted central value is also closer to the data.

\section{Conclusions}\label{sec5}

In this study we calculated the total and differential cross-sections and cross-section ratios using the MadGraph up to $O(\alpha_s^2)$ with a massive charm quark $m_c = 1.55$ GeV for three NNLO PDF sets: MSTW2008, CT10, and CT14NNLO. Subsequently, we compared the experimental measurements of $W+$ charm production at $\sqrt{s}=7$ TeV at LHC.
In our calculation, we use the same kinematic cuts as experimental measurements: $p_T^{jet}>25$ GeV, $|\eta^{jet}|<2.5$, and  $|\eta_l|<2.1$ GeV, and two different transverse momentum cuts $p_T^l > 25$GeV for the $W \to \mu\nu$ channel and $p_T^l>35$ GeV for the $W \to \mu\nu$ and $W \to e\nu$ channels. In our calculation, both the factorization and the renormalization scales are set to the value of the $W$ boson mass, and $\alpha_s(M_Z)$ is set to the central value provided by the respective PDF groups.
Our results are  summarized in Tables \ref{Totxsectab} - \ref{diffratiotab} and in Figures \ref{Totxsecfig} - \ref{diffratioxsecfig},
%\ref{Totxsectab},\ref{LOcon},\ref{diffxsectab},\ref{normdiffxsectab},\ref{totalratiotab},\ref{difratiotab}
 where the central value of the prediction and the PDF uncertainty are given. The theoretical predictions from various PDFs agree well with  experimental measurements. However, there are some differences depending on the PDFs  used in the calculations. For example, unlike the assumption in MSTW20018 NNLO PDFs, the CT10 and CT14  assume $s=\bar s$ in the proton,  yielding to a total and differential cross-sections  ratio dominated by the $d-\bar d$ asymmetry.
 The total and differential cross-sections are larger for the $W^- + c$  production than for $W^+ + c$, because the former process involves a $d$,   whereas the latter involves $\bar d$  (sea) antiquark. 
Hence, both total and differential cross-section ratios are smaller than $1.0$.

 Fig. \ref{pdf:cor} shows that the observable from the CMS 7TeV $W+$ charm production has a strong correlation with the strange(anti) quark PDFs, therefore these measurements also provide a direct constraint on the strange(anti) quark content of the proton.

Furthermore, using the ePump updating method, and CMS 7TeV $W+$ charm production data  at lepton transverse momentum $p_T^l>35$ GeV, we find that these data sets mainly reduce the $s(x,Q)$ PDF error band and increase magnitude of its best fit in the $x<0.4$ region for both $Q=1.3$ GeV and
$Q=100$ GeV. In Fig.\ref{utheo}, we also compare the predictions from  CT14NNLO PDFs and  CT14NNLO+Wc PDFs.

\section*{Acknowledgments}
We thank Tie-Jiun Hou many helpful discussions. This work is supported by the National Natural Science Foundation of China under the Grant No. 11965020.

%\section*{Reference}

\end{document}